\documentclass[a4paper,11pt]{article}
\usepackage{jheppub} 
\usepackage{lineno}
\allowdisplaybreaks
\usepackage[T1]{fontenc}   
\usepackage[utf8]{inputenc} 

\title{$T_7$ Flavour Symmetry gym: The Key to Unlock the Neutrino Mass Puzzle}
\author[a]{Muhammad Waheed Aslam,}
\author[a]{Abrar Ahmad Zafar,}

\affiliation[a]{Department of Physics, University of the Punjab,\\
Lahore, Pakistan}

\emailAdd{waheed-979531@pu.edu.pk}

\abstract{Recent research has indicated that the Standard Model (SM), while historically highly effective, is found to be insufficient due to its prediction of zero mass for neutrinos. With the exception of a few, the majority of the parameters related to neutrinos have been determined by neutrino oscillation experiments with excellent precision. Experiments on neutrino oscillation and neutrino mixing have shown that neutrinos are massive. To fill in gaps, discrete symmetries are becoming more common alongside continuous symmetries while describing the observed pattern of neutrino mixing. Here, we present a $T_7$ flavor symmetry to explain the masses of charged leptons and neutrinos. The light neutrino mass matrix is derived using seesaw mechanism of type I, which involves the Dirac neutrino mass matrix as well as the right-handed neutrino mass matrix. For normal and inverted mass hierarchy, we estimate the Pontecorvo-Maki-Nakagawa-Sakata matrix ($U_{PMNS}$), three mixing angles, $\theta_{12}$, $\theta_{23}$ and $\theta_{13}$, the masses of three neutrinos, effective Majorana neutrino mass parameter $\langle m_{ee} \rangle$ and the other model parameters by using a powerful meta-heuristic and population-based optimization algorithm, i.e. particle swarm optimization (PSO).}

\begin{document}
\maketitle
\flushbottom

\section{Introduction}
\label{sec:intro}

In order to justify missing energy in beta decay, 
Wolfgang Pauli suggested the presence of particles called 
neutrinos in 1930. Antineutrinos 
produced in a nuclear reactor studies at Savannah River in South 
Carolina, USA, were discovered in 1956. Through detection of solar 
neutrinos at an underground experiment carried out by an 
American scientist Raymond Davis Jr. 
\cite{cleveland1995update}, an earliest indication was obtained that neutrinos had mass. The phenomenon of neutrino 
oscillation (ability of neutrino to change its flavor while 
travelling) was first introduced by Pontecorvo in 1957 
\cite{pontecorvo1958mesonium}. The Japanese experiment 
Super-Kamiokande (SK) first time observed this phenomenon in 1998
by measuring that during their journey through the Earth, the 
number of muon neutrinos decreases to half at the exit as compared to incoming number. This decrease was thought to be due to conversion of muon neutrino into tau neutrinos \cite{kamiokande1998evidence}.

In 2002, the SNO (Sudbury Neutrino Observatory) verified the flavour conversion in solar neutrinos \cite{kamiokande1998evidence}. The results of solar neutrino oscillations were further verified; by studying artificial neutrinos produced at nuclear reactors at KamLAND in Japan \cite{eguchi2003first}, utilizing neutrino beams fired over hundreds of kilometres like those at the T2K experiment in Japan \cite{ahn2006measurement}, by the Fermilab-MINOS experiment in United States \cite{michael2006observation} and by CERN-OPERA experiment in Europe \cite{sh2009opera}. This finding sparked a flurry of research into the mechanisms by which neutrinos acquire mass.

 In order to connect the SM neutrino states ($\nu_e$, $\nu_{\mu}$, $\nu_{\tau}$) to the neutrino mass states ($v_1$, $v_2$, $v_3$), the lepton mixing matrix U is used to describe the coupling strength as \cite{maki1962remarks,lee1977muon}
\begin{equation}
\label{eq:e1}
\begin{aligned}
\begin{pmatrix}
\nu_e \\
\nu_\mu \\
\nu_\tau 
\end{pmatrix} = 
\begin{pmatrix}
U_{e1} & U_{e2} & U_{e3}\\
U_{\mu1} & U_{\mu2} & U_{\mu3}\\
U_{\tau1} & U_{\tau2} & U_{\tau3}
\end{pmatrix}  \begin{pmatrix}
v_1 \\
v_2 \\
v_3 
  \end{pmatrix}
\end{aligned}.
\end{equation}
In mass state 2, the electron neutrino, muon neutrino and tau 
neutrino contain roughly equal probabilities, 
$|U_{e2}|\approx|U_{\mu2}|\approx|U_{\tau2}|\approx1/\sqrt{2}$, 
(trimaximal mixing). In mass state 3, the muon neutrino and tau 
neutrino contain roughly equal probabilities, 
$|U_{\mu3}|\approx|U_{\tau3}|\approx1/\sqrt{2}$
and the probability of electron neutrino is negligible, 
$|U_{e3}|\approx0$ (maximal mixing) . Many models have been constructed that are in 
agreement with $|U_{e3}|\approx 0$. But in reference \cite{apollonio1999limits}, the
probability of electron neutrino could only contain a very small 
amount, $|U_{e3}|<0.2$. Collaborations; Daya Bay \cite{an2012observation,an2015new}, RENO \cite{a2012observation,choi2016observation} and Double Chooz \cite{abe2012indication,abe2014improved}, have measured $|U_{e3}|\approx 
0.15$. Because this mixing matrix element is non-zero, it restricts 
a number of previously proposed simple mixing patterns and models, 
leading to a number of new discoveries. According to the global fit 
\cite{king2013neutrino}, the mixing angles are $\theta_{12}$ (solar neutrino mixing 
angle)$=34^\circ \pm 1^\circ$, $\theta_{23}$ (atmospheric neutrino
mixing angle) $=42^\circ \pm 3^\circ$, $\theta_{13}$ (reactor 
angle)$=8.5^\circ\pm 0.5^\circ$. Neutrino oscillations wouldn't 
violate the $CP$ if the reactor angle was zero \cite{king2013neutrino}. We can no 
longer dismiss the existence of phases due to the measurement of the
reactor angle.

When the phases are taken into account, $U_{PMNS}$ matrix may be 
parameterized in terms of three mixing angles ($\theta_{12}$, 
$\theta_{23}$ and $\theta_{13}$ representing the solar, atmospheric, 
and reactor angles, respectively.) with a Dirac phase for $CP$ 
violation $\delta$ lying in the range $[0, 2\pi]$, and two $CP$ 
violation Majorana phases $\beta_i$ $(i=1,\ 2)$ as shown below \cite{vien2016delta}.
\begin{equation}
\label{eq:e2}
\begin{aligned}
U_{PMNS}= 
  \begin{pmatrix}
    c_{12}c_{13} & s_{12}c_{13} & s_{13} e^{i\delta}\\
    -s_{12}c_{23}-c_{12}s_{13}s_{23}e^{i\delta} & c_{12}c_{23}-s_{12}s_{13}s_{23}e^{i\delta} & c_{13}s_{23}\\
    s_{12}s_{23}-c_{12}c_{23}s_{13}e^{i\delta} & -c_{12}s_{23}-s_{12}s_{13}c_{23}e^{i\delta} & c_{13}c_{23}
  \end{pmatrix} P_{12}
\end{aligned}
\end{equation}
with, $c_{ij}=\cos{\theta_{ij}}$, $s_{ij}=\sin{\theta_{ij}}$ and 
$P_{12}=\begin{pmatrix}
    1 & 0 & 0\\
    0 & e^{i\beta_{1}} & 0\\
    0 & 0 & e^{i\beta_{2}}
\end{pmatrix}.$

In spite of the findings mentioned above, 
neutrinos remain least understood.  Only five parameters (three mixing 
angles and two mass square differences) out of seven parameters related with neutrino mass and mixing are measured so far and the $CP$ violating phases $\delta$ 
and $\beta_{1}$ and $\beta_{2}$ have not been measured 
yet. Therefore, nature of neutrino (Dirac or 
Majorana) is still unknown.

We also don’t know still about the lightest 
neutrino 
mass. But the upper limit of $m_{\nu}<0.8$ $eV$ at 90\% 
 CL was obtained in 2022 by the KATRIN Collaboration \cite{katrin2022direct}. To investigate physics that goes outside the 
Standard Model, such sterile neutrinos, with the 
KATRIN experiment, high-precision $\beta$-
spectroscopy has been demonstrated to be a powerful
tool for probing the neutrino mass with extraordinary 
sensitivity. The future KATRIN data will be crucial 
in determining the neutrino-mass parameters together 
with cosmic probes and searches for neutrinoless 
double-decay \cite{katrin2022direct}. In \cite{particle2022review}, upper and lower bound 
limits of $\Sigma m$ are $0.12$ $eV$ and $0.06$ $eV$ 
respectively. In \cite{de20212020}, the upper bound limit for 
normal hierarchy is $\Sigma m<0.12$ $eV$ and for 
inverted mass hierarchy is $\Sigma m<0.15$ $eV$.

Neutrino oscillations are only sensitive to the mass 
square differences, and are not sensitive to the 
absolute mass, so absolute masses of neutrinos are 
still unknown. Therefore, we have two orderings of 
square masses; normal mass hierarchy 
$(m_3^2>m_2^2>m_1^2)$ and inverted mass hierarchy 
$(m_2^2>m_1^2>m_3^2)$, since recent neutrino 
experiments have not yet specified the ordering of 
square masses. experiments determine square neutrinos 
mass differences only \( m_2^2-m_1^2\sim 7.49\times 10
^{-5} {eV}^2\) and \(m_3^2-m_2^2\sim2.513\times10^{-3} 
{eV}^2 (m_3^2-m_2^2\sim-2.510\times10^{-3} {eV}
^2) \) for normal (inverted) mass hierarchy \cite{esteban2024nufit} and 
provide no details regarding the absolute values.

The seesaw mechanism \cite{aguila2019inverse} stands out as
a well-known method for probing  tiny neutrino masses. 
However, given their incredibly large mass scale, the 
absence of experimental evidence for right-handed neutrinos 
continues to be a major obstacle. It is necessary to 
introduce new physics at TeV scale suitable with 
the linear seesaw mechanism \cite{mohapatra1986mechanism,hirsch2009a4,hernandez2018variant,carcamo2021controlled} in order to
resolve this problem; this might be investigated 
through experiments at the Large Hadron Collider 
(LHC).

Neutrino masses with discrete symmetries have been a topic of interest in particle physics since the discovery of neutrino oscillations in the late 20th century. The missing neutrino parameter should be predicted theoretically, when experiments do not provide sufficient information. The SM also does not provide the true information about the masses of neutrinos. The possibility of incorporating discrete symmetries into models of neutrino mass generation, has garnered attention as a potential means of explaining the small but non-zero masses of neutrinos \cite{hirsch2009a4,hernandez2018variant,carcamo2021controlled,sruthilaya2018a_4,borah2019linear,hernandez2023linear}. $T_7$ is a specific type of discrete symmetry that can be employed to explain the observed patterns in neutrino mass and mixing. Researchers have explored its applications in various theoretical models to account for the peculiar properties of neutrinos. Key papers and sources for further exploration include \cite{T71,T72,T73,T74,T75,T76,T77,T78,T79,T710,T711,T712,T713}. In these studies, for the purpose of calculating the masses and mixing of quarks and/or leptons, the SM Higgs has already been used as a $T_7$ singlet. In our study, the SM Higgs is taken as the $T_7$ triplet.

While recent developments have made progress possible, analytical evaluation in such models continues to be a significant challenge because of the nonlinear and complex nature of equations involved, especially ones concerning neutrino mass eigenstates and mixing matrices. This complexity commonly requires one to use advanced numerical optimization methods. The particle swarm optimization (PSO) is a population-based optimization algorithm drawing inspiration from biological swarms' collective movement and foraging behaviors, e.g., flocks of birds or schooling fish. Originally formulated in 1995 by Russell Eberhart and James Kennedy, PSO borrows conceptual ideas from genetic algorithms (GAs) but lays out a unique method to search for and exploit space \cite{panda2008comparison, okwu2021particle, yu2020comparison}. PSO consists of a population of possible solutions—known as particles—searching for a route through solution space by modifying their position according to local and global knowledge, arriving at optimal or near-optimal solutions. Owing to its flexibility and efficiency, PSO has found extensive applications across a variety of domains. These include engineering design optimization \cite{guedria2016improved, kaveh2013engineering, kumar2021design}, image processing tasks \cite{djemame2019solving, singh2014image, pramanik2015image}, financial modeling and forecasting \cite{azayite2019financial, chiam2009memetic, pan2022design, marinakis2009ant}, and the training of artificial neural networks \cite{meissner2006optimized, rauf2018training}. Its utility has also been demonstrated in high-dimensional data clustering \cite{esmin2015review, rana2011review}, chaotic system parameter estimation \cite{he2007parameter, alatas2009chaos}, optimization of core loading configurations in nuclear reactors \cite{babazadeh2009optimization}, and solving complex physical problems such as nonlinear reference frame alignment \cite{ibrahim2019hybridization} and optimal reactive power distribution \cite{subbaraj2010hybrid}. Moreover, PSO has proven to be highly effective in estimating parameters in diverse scientific models, including the analysis of optical properties in multilayer thin films \cite{jiang2020multilayer, yue2019determination, rabady2014global, ruan2016determination} and autoregressive moving average (ARMA) models \cite{mehmood2019nature, yetis2014forecasting, yassin2016binary}. It has also been applied in the estimation of parameters for electromagnetic wave models \cite{akbar2019novel}. Particle Swarm Optimization (PSO) and other optimization algorithms have been previously used in our recent work to analyze neutrino masses and mixing patterns within discrete symmetry models \cite{aslam2025particle, Aslam:2025vvp, doi:10.1142/S0217732325501068}. The algorithm's popularity can be attributed to its straightforward implementation, low memory requirements, and robust performance in handling non-linear, high-dimensional, and multimodal objective functions \cite{stacey2003particle, eberhart1995new, toushmalani2013gravity}. 

Unlike earlier models based on the $T_7$ flavor symmetry, the present work adopts a novel computational approach by employing Particle Swarm Optimization (PSO), a powerful, general-purpose metaheuristic algorithm that is widely used for solving complex optimization problems. This technique has been used in this study to optimize the multidimensional parameter space of the model, allowing for precise predictions for the masses of the three light neutrinos, the leptonic mixing matrix elements, the phases of the Dirac and Majorana CP-violation, and other phenomenologically important parameters. These are tightly constrained by the existing experimental information from the studies of the oscillation and mixing of neutrinos. Furthermore, the PSO technique has been extended to optimize of the flavon sector, and we have been able to estimate the masses of the flavon fields. Interestingly, these flavon masses lie within a range that may be accessible to future experimental verification, thereby enhancing the testability and predictive power of the proposed model.

The structure of our paper unfolds as follows: In the subsequent section \ref{sec:The Model}, we provide our model based on $T_7$ non-abelian discrete symmetry. Subsection \ref{sec:Char lept} is dedicated to demonstrate the superpotential terms for charged leptons and demonstrate the natural explanation of the hierarchical pattern in charged lepton masses. Moving on to subsection \ref{sec:Neutrinos masses}, we delve into the realm of neutrino masses and mixing, showcasing how our model successfully realizes the observed mixing pattern. Section \ref{sec:Numericsal Analysis} is dedicated to the PSO technique, its working principle, and numerical analysis of the model parameters using PSO and derives the $U_{PMNS}$, mixing angles, and neutrino masses. Section \ref{sec:Scalar} presents the $T_7$ invariant scalar potential and conditions for minimization of the potential and derives the possible vacuum alignments for the vacuum expectation values (VEVs) of the scalars. Finally, we summarize our findings and present our conclusions in 
section \ref{sec:Conclusion}. In appendix \ref{sec:App}, we provide an intricate description of the $T_7$ group. 
\section{The Model}
\label{sec:The Model}
In this paper, we suggest a model which is based on $T_7$ non abelian discrete symmetry in which lepton doublet is taken as $T_7$ triplet $\overline{\textbf{3}}$ and right-handed neutrinos, $\nu_{R_{\alpha}}$ as singlets, $\textbf{1}_\alpha$, while right-handed charged lepton fields $l^c_{e,\mu,\tau}$ are taken as different singlet representations of $T_7$. We also consider SM Higgs scalar (h) and following scalars: $\phi_l$, $\phi_\nu$, $\xi_1$, $\xi_2$ and $\xi_3$, whose representations are given in table \ref{tab:charge assignments}, that break the flavor symmetry.
In the scalar sector, we introduced five extra falvon fields, two of them transforming under triplet of $T_7$ and other three are $T_7$ singlets in our model which ensure natural hierarchy among lepton masses, the VEVs of these fields break the flavor symmetry and these are responsible to acquire the masses for leptons. Such type of flavon fields are also utilized in numerous articles, including: \cite{chauhan2024phenomenology, barry2011light,adhikari2016new, borah2017nonzero, das2019active, sarma2019compatibility, krishnan2020realization, vien2022b, miskaoui2024neutrino}. We also introduce $Z_3$ symmetry to constrain all unwanted terms. Moreover, We have only one $SU(2)$ Higgs doublet “h” which transformed under $T_7$ as triplets. $\phi_l$ and $\phi_\nu$ are $SU(2)$ singlets which are transformed under $T_7$ triplets. So we have no any flavor changing neutral currents (FCNCs).  In our model we have only one Higgs doublet is used to break the electroweak symmetry, and SM singles only are involved in spontaneous breaking of $T_7$, so our model is FCNC free.  
\begin{table}[htbp]
\centering
\begin{tabular}{llllllllllll}
\hline
 Fields& $\bar{L}$($l_e$, $l_\mu$, $l\tau$) & $l_e^c$ &$l_\mu^c$  &$l_\tau^c$  & $\nu_{R_{\alpha}}$ & h &$\phi_l$ & $\phi_\nu$ &  $\xi_1$&  $\xi_2$ &$\xi_3$ 
 \\
\hline
$SU(2)$ & $2$ & $1$&  $1$&  $1$ & $1$ & $2$ & $1$ &$1$ &$1$ & $1$& $1$\\
\hline
$T_7$ & $\overline{\textbf{3}}$ & $\textbf{1}_0$&  $\textbf{1}_2$&  $\textbf{1}_1$&$\textbf{1}_\alpha$& \textbf{3}&\textbf{3} &$\overline{\textbf{3}}$ &$\textbf{1}_0$ & $\textbf{1}_1$& $\textbf{1}_2$\\
\hline
$Z_3$ & $\omega$ & $\omega^2$&  $\omega^2$&  $\omega^2$ &$\omega$ & $1$ & $1$ &$\omega$ &$\omega$ & $\omega$& $\omega$\\
\hline
\end{tabular}
\caption{The representations of $T_7$ for fermions and scalars\label{tab:charge assignments}}
\end{table}
\subsection{Charged lepton masses and mixing}
\label{sec:Char lept}
The Yukawa interaction for charged leptons is given as 
\begin{equation}
\label{eq:e3}
\begin{aligned}
  -\mathcal{L}_l=\frac{y^e}{\Lambda} \phi_l\bar{L}hl_e^c+\frac{y^{\mu}}{\Lambda}  \phi_l\bar{L}hl_{\mu}^c+\frac{y^{\tau} }{\Lambda}\phi_l\bar{L}hl_{\tau}^c+H.c.,
\end{aligned}
\end{equation}
where, $y^e$, $y^{\mu}$, $y^{\tau}$ are Yukawa couplings and $\Lambda$ is cutoff scale. Due to the VEVs (see section \ref{sec:Scalar}), the symmetry breaking occurs and generates charged lepton mass matrix,
\begin{equation}
\label{eq:e4}
\begin{aligned}
M_{l}=\frac{H}{\Lambda} 
  \begin{pmatrix}
    y^{e}\nu_{3} & \omega y^{\mu}\nu_{3} & \omega^2 y^{\tau}\nu_{3}\\
       y^{e}\nu_1 & y^{\mu}\nu_1 & y^{\tau}\nu_1\\
       y^{e}\nu_{2} & \omega^2 y^{\mu}\nu_{2} & \omega y^{\tau}\nu_{2}\\
  \end{pmatrix}
\end{aligned}.
\end{equation}
The lepton mixing matrix, which describes the mixing pattern between 
neutrinos of different flavors, can deviate slightly from a specific 
form known as the tri-bimaximal mixing pattern. In the tri-bimaximal mixing pattern, the mixing matrix has a specific structure that involves the angles of $45^\circ$, $35.3^\circ$, and 
$0^\circ$ (approximately). However, experimental data has shown that 
there are small deviations from this idealized pattern. So here we 
use a way to account for these deviations by introducing small 
parameters $\eta_2$ and $\eta_3$. These parameters are assumed to be less than 1
in magnitude, indicating the magnitude of the deviations from the tri-
bimaximal form. In this idea, the lepton mixing matrix is described 
in terms of a common parameter "$\nu$" as well as the parameters $\eta_2$ and $\eta_3$. The components of mixing matrix 
involving $\nu_2$ and $\nu_3$ are adjusted to account for these minor 
deviations from the tri-bimaximal form. Specifically: $\nu_1$ is 
chosen as the common parameter $\nu$, $\nu_2$ is changed to 
$(1+\eta_2)\nu$ and $\nu_3$ is changed to $(1+\eta_3)\nu$. 
These deviations are responsible for the departure of the actual 
lepton mixing pattern from the idealized one. This method is often 
used in neutrino physics to explain little variations from theoretically predicted results in lepton mixing matrices \cite{vien2016delta}. The perturbations \(\eta_{2,3}\) introduced in the charged lepton sector originate from deviations in the VEVs alignment of the flavon fields. In 
short, it enables the use of experimental data to provide a more 
precise explanation of neutrino oscillation processes. By using this
concept, the charged lepton mass matrix becomes
\begin{equation}
\label{eq:e5}
\begin{aligned}
M_{l}=\frac{H \nu}{\Lambda} 
  \begin{pmatrix}
    y^{e}(1+\eta_3) & \omega y^{\mu}(1+\eta_3) & \omega^2 y^{\tau}(1+\eta_3)\\
       y^{e} & y^{\mu} & y^{\tau}\\
       y^{e}(1+\eta_2) & \omega^2 y^{\mu}(1+\eta_2) & \omega y^{\tau}(1+\eta_2)\\
  \end{pmatrix}
\end{aligned}.
\end{equation}
Now, defining two unitary matrices:
\begin{equation}
\label{eq:e6}
\begin{aligned}
U_{OL}=\frac{1}{\sqrt{3}} 
  \begin{pmatrix}
    1 & 1 & 1\\
       \omega & 1 & \omega^2\\
      \omega^2 & 1& \omega
  \end{pmatrix},\qquad
  U^{\prime}_{OL}=\frac{1}{\sqrt{3}} 
  \begin{pmatrix}
    1 & 1 & 1\\
       \omega^2 & 1 & \omega\\
      \omega & 1& \omega^2
  \end{pmatrix},
\end{aligned}
\end{equation}
and by using unitary matrix given in equation \ref{eq:e6}, the charged lepton mass matrix is given as follows
\begin{equation}
\label{eq:e7}
\begin{aligned}
M_L=U^{\prime}_{OL}.M_l=\frac{H\nu}{\sqrt{3}\Lambda}
  \begin{pmatrix}
    1 & 1 & 1\\
       \omega^2 & 1 & \omega\\
      \omega & 1& \omega^2
  \end{pmatrix}
    \begin{pmatrix}
    y^{e}(1+\eta_2) & \omega^2 y^{\mu}(1+\eta_2) & \omega y^{\tau}(1+\eta_2)\\
       y^{e} & y^{\mu} & y^{\tau}\\
       y^{e}(1+\eta_3) & \omega y^{\mu}(1+\eta_3) & \omega^2 y^{\tau}(1+\eta_3)\\
  \end{pmatrix}.
\end{aligned}
\end{equation}

Entities of diagonalized charged lepton mass matrix are calculated as $U^{\dagger}_{L} M_L U_R=diag(m_1,m_2,m_3)$, where $U_L$ and $U_R$ are given in equation \ref{eq:e42}.
\begin{equation}
\label{eq:e8}
\begin{aligned}
m_1=\frac{3\sqrt{3} H\nu y^e (1+\eta_2)(1+\eta_3)}{\Lambda(3+2\eta_3+\eta_2 (2+\eta_3))},\\
m_2=\frac{3\sqrt{3} H\nu y^\mu (1+\eta_2)(1+\eta_3)}{\Lambda(3+2\eta_3+\eta_2 (2+\eta_3))},\\
m_3=\frac{3\sqrt{3} H\nu y^\tau (1+\eta_2)(1+\eta_3)}{\Lambda(3+2\eta_3+\eta_2 (2+\eta_3))},
\end{aligned}
\end{equation}
The VEVs of fields and values of $\eta_2$ and $\eta_3$ are given in section \ref{sec:Numericsal Analysis}. The experimental values of masses of charged leptons at the weak scale are given as \cite{particle2022review} $m_e=0.511$ $MeV$, $m_\mu=105.658$ $MeV$ and $m_\tau=1776.86$ $MeV$. In charged lepton sector, on the behalf of VEVs of fields and values of $\eta_2$ and $\eta_3$, the Yukawa couplings follows that $y^e<<y^\mu<<y^\tau$, and ratios of Yukawa couplings are obtained as,
$\frac{y^e}{y^\mu} =0.0048$, $\frac{y^e}{y^\tau} =0.00029$, $\frac{y^\mu}{y^\tau}=0.0595$.
\subsection{Neutrinos masses}
\label{sec:Neutrinos masses}
By considering the lepton doublet, $l_i$, as a $T_7$ triplet and the right-handed neutrinos, $\nu_{R_\alpha}$, as singlets, we can formulate the following invariant interaction terms:
\begin{equation}
\label{eq:e12}
\begin{aligned}
-\mathcal{L}_D=\frac{\lambda_1}{\Lambda}(\bar{L} \phi_\nu h)_{\textbf{1}_0}(\nu_{R_1})_{1_0}+\frac{\lambda_2}{\Lambda}(\bar{L} \phi_\nu h)_{\textbf{1}_1}(\nu_{R_2})_{1_2}+\frac{\lambda_3}{\Lambda}(\bar{L} \phi_\nu h)_{\textbf{1}_2}(\nu_{R_3})_{1_1}+H.c.,
\end{aligned}
\end{equation}
after expanding equation \ref{eq:e12} and by using multiplication rules given in appendix \ref{sec:App}, one can get following  
\begin{equation}
\label{eq:e13}
\begin{aligned}
-\mathcal{L}_D=&\frac{1}{\Lambda}\lambda_1 (h_2 \bar{L}_1 \phi_{\nu_1}+h_3 \bar{L}_2 \phi_{\nu_2}+h_1 \bar{L}_3 \phi_{\nu_3})\nu_{R_1}\\
&+\frac{1}{\Lambda}\lambda_2 (h_2 \bar{L}_1 \phi_{\nu_1}+\omega^2h_3 \bar{L}_2 \phi_{\nu_2}+\omega h_1 \bar{L}_3 \phi_{\nu_3})\nu_{R_2}\\
&+\frac{1}{\Lambda}\lambda_3 (h_2 \bar{L}_1 \phi_{\nu_1}+\omega h_3 \bar{L}_2 \phi_{\nu_2}+\omega^2 h_1 \bar{L}_3 \phi_{\nu_3})\nu_{R_3}
\end{aligned}
\end{equation}
and due to the VEVs of $\phi_\nu$ and $h$ (see section \ref{sec:Scalar}), we get 
\begin{equation}
\label{eq:e14}
\begin{aligned}
M_D=\frac{H\chi}{\Lambda}
  \begin{pmatrix}
    \lambda_1 & \lambda_2 & \lambda_3\\
       \lambda_1 & \omega^2\lambda_2 & \omega\lambda_3\\
      \lambda_1 & \omega\lambda_2& \omega^2\lambda_3
  \end{pmatrix}.
\end{aligned}
\end{equation}
Note: all Dirac neutrino masses are derived from identical non-renormalizable interactions at the scale $H\chi/\Lambda$. As a result, there should be no anticipation of any hierarchy among these masses.

Additionally, Majorana mass terms for $\nu_R$ can be derived through renormalizable interactions involving the invariant interactions under $T_7$ between right-handed neutrinos and the $T_7$ singlets $\xi_i$:
\begin{equation}
\label{eq:e15}
\begin{aligned}
\mathcal{L}_M=f_{ijk}\overline{\nu}^{c}_{R_{1_j}}\nu_{R_{1_j}}\xi_k,
\end{aligned}
\end{equation}
due to the multiplication rules for singlet representations under $T_7$, the invariants responsible for generating masses for right-handed neutrinos are:
\begin{equation}
\label{eq:e16}
\begin{aligned}
f_{111}\overline{\nu}^{c}_{R_{1_0}}\nu_{R_{1_0}}\xi_1,\qquad  f_{222}\overline{\nu}^{c}_{R_{1_1}}\nu_{R_{1_1}}\xi_2,\qquad   
f_{333}\overline{\nu}^{c}_{R_{1_2}}\nu_{R_{1_2}}\xi_3,\\
f_{123}\overline{\nu}^{c}_{R_{1_0}}\nu_{R_{1_1}}\xi_3,\qquad 
f_{132}\overline{\nu}^{c}_{R_{1_0}}\nu_{R_{1_2}}\xi_2,\qquad 
f_{213}\overline{\nu}^{c}_{R_{1_1}}\nu_{R_{1_0}}\xi_3,\\
f_{231}\overline{\nu}^{c}_{R_{1_1}}\nu_{R_{1_2}}\xi_1,\qquad 
f_{313}\overline{\nu}^{c}_{R_{1_2}}\nu_{R_{1_0}}\xi_2,\qquad 
f_{321}\overline{\nu}^{c}_{R_{1_2}}\nu_{R_{1_1}}\xi_1.
\end{aligned}
\end{equation}
where, $f_1\equiv f_{111}$, $f_2\equiv f_{123}=f_{213}$, $f_3\equiv f_{132}=f_{313}$, $f_4\equiv f_{222}$, $f_5\equiv f_{231}=f_{321}$, $f_6\equiv f_{333}$ and due to VEVs of $\xi_1$, $\xi_2$ and $\xi_3$ (see section \ref{sec:Scalar}), we get
\begin{equation}
\label{eq:e18}
\begin{aligned}
M_R=
  \begin{pmatrix}
    f_{1} u_1 & f_{2} u & f_{3} u\\
       f_{2} u & f_{4} u & f_{5} u_1\\
      f_{3} u & f_{5} u_1& f_{6} u
  \end{pmatrix},
\end{aligned}
\end{equation}
Typically, the light neutrino mass matrix is derived using seesaw mechanism of type I, which involves the Dirac neutrino mass matrix as well as the right-handed neutrino mass matrix as
\begin{equation}
\label{eq:e19}
\begin{aligned}
M^{\prime}_\nu&=-m_D M_R^{-1} m_D^{T}= \frac{1}{V}\begin{pmatrix}
    P & Q & R\\
       Q& S & T\\
      R & T & U
  \end{pmatrix},
\end{aligned}
\end{equation}
with,
\begin{equation}
\label{eq:pa}
\begin{aligned}
V=\Lambda ^2 \left(\left(f_6 f_2^2+f_3^2 f_4\right) u^3-\left(2 f_2 f_3 f_5+f_1 f_4 f_6\right) u_1 u^2+f_1 f_5^2 u_1^3\right),
\end{aligned}
\end{equation}
\begin{equation}
\label{eq:para1}
\begin{aligned}
P =& H^2 \chi^2 \big(-f_5^2 \lambda_1^2 u_1^2 + 2f_5 u_1 \big(f_3 \lambda_1 \lambda_2 u + \lambda_3 \big(f_2 \lambda_1 u - f_1 \lambda_2 u_1\big)\big) + f_4 u \big(f_6 \lambda_1^2 u \\&+ \lambda_3 \big(f_1 \lambda_3 u_1 - 2f_3 \lambda_1 u\big)\big) - u \big(\lambda_2^2 \big(f_3^2 u - f_1 f_6 u_1\big) + 2f_2 \lambda_2 u \big(f_6 \lambda_1 - f_3 \lambda_3\big) + f_2^2 \lambda_3^2 u\big)\big),
\end{aligned}
\end{equation}
\begin{equation}
\label{eq:para2}
\begin{aligned}
Q =& \frac{1}{2} \bigg[ H^2 \chi^2 \big( -f_5^2 \lambda_1^2 u_1^2 + 2f_5 u_1 \big( f_3 \lambda_1 \lambda_2 u + \lambda_3 \big( f_2 \lambda_1 u - f_1 \lambda_2 u_1 \big) \big) + f_4 u \big( f_6 \lambda_1^2 u + \\&\lambda_3 \big( f_1 \lambda_3 u_1 - 2f_3 \lambda_1 u \big) \big) - u \big( \lambda_2^2 \big( f_3^2 u - f_1 f_6 u_1 \big) + 2f_2 \lambda_2 u \big( f_6 \lambda_1 - f_3 \lambda_3 \big) + f_2^2 \lambda_3^2 u \big) \big) \bigg],
\end{aligned}
\end{equation}
\begin{equation}
\label{eq:para3}
\begin{aligned}
R = &\frac{1}{2} \bigg[ H^2 \chi^2 \big( \lambda_2 \big( 2\lambda_3 \big( f_1 f_5 u_1^2 - f_2 f_3 u^2 \big) + (-1 - i\sqrt{3}) \lambda_1 u \big( f_2 f_6 u - f_3 f_5 u_1 \big) \big) \\&+ (1 - i\sqrt{3}) \lambda_2^2 u \big( f_3^2 u - f_1 f_6 u_1 \big) - \big( f_5 \lambda_1 u_1 - f_2 \lambda_3 u \big) \big( 2f_5 \lambda_1 u_1 + (1 + i\sqrt{3}) f_2 \lambda_3 u \big) \\&+ f_4 u \big( 2f_6 \lambda_1^2 u + i\lambda_3 \big( (\sqrt{3} + i) f_3 \lambda_1 u - (\sqrt{3} - i) f_1 \lambda_3 u_1 \big) \big) \big) \bigg],
\end{aligned}
\end{equation}
\begin{equation}
\label{eq:para4}
\begin{aligned}
S =& \frac{1}{2} \bigg[ H^2 \chi^2 \big( -2f_5^2 \lambda_1^2 u_1^2 + u \big( (1 - i\sqrt{3}) \lambda_2^2 \big( f_3^2 u - f_1 f_6 u_1 \big) + 2f_2 \lambda_2 u \big( 2f_3 \lambda_3  \\&+ (1 + i\sqrt{3}) f_6 \lambda_1 \big)+ (1 + i\sqrt{3}) f_2^2 \lambda_3^2 u \big) + 2f_5 u_1 \big( i(\sqrt{3} + i) f_2 \lambda_1 \lambda_3 u - \lambda_2 \big( 2f_1 \lambda_3 u_1\\& + (1 + i\sqrt{3}) f_3 \lambda_1 u \big) \big) + f_4 u \big( 2f_6 \lambda_1^2 u + \lambda_3 \big( (2 - 2i\sqrt{3}) f_3 \lambda_1 u + (-1 - i\sqrt{3}) f_1 \lambda_3 u_1 \big) \big) \big) \bigg],
\end{aligned}
\end{equation}
\begin{equation}
\label{eq:para5}
\begin{aligned}
T =& H^2 \chi^2 \big( -f_5^2 \lambda_1^2 u_1^2 + f_5 u_1 \big( \lambda_3 \big( f_1 \lambda_2 u_1 - f_2 \lambda_1 u \big) - f_3 \lambda_1 \lambda_2 u \big) - u \big( \lambda_2^2 \big( f_3^2 u - f_1 f_6 u_1 \big) \\&+ f_2 \lambda_2 u \big( f_3 \lambda_3 - f_6 \lambda_1 \big) + f_2^2 \lambda_3^2 u \big) + f_4 u \big( f_6 \lambda_1^2 u + \lambda_3 \big( f_3 \lambda_1 u + f_1 \lambda_3 u_1 \big) \big) \big),
\end{aligned}
\end{equation}
\begin{equation}
\label{eq:para6}
\begin{aligned}
U =& \frac{1}{2} \bigg[ H^2 \chi^2 \big( -2f_5^2 \lambda_1^2 u_1^2 + 2f_5 u_1 \big( i(\sqrt{3} + i) f_3 \lambda_1 \lambda_2 u + \lambda_3 \big( -2f_1 \lambda_2 u_1 + \\&(-1 - i\sqrt{3}) f_2 \lambda_1 u \big) \big) + u \big( (1 + i\sqrt{3}) \lambda_2^2 \big( f_3^2 u - f_1 f_6 u_1 \big) + 2f_2 \lambda_2 u \big( 2f_3 \lambda_3 + (1 - i\sqrt{3}) f_6 \lambda_1 \big) \\&+ (1 - i\sqrt{3}) f_2^2 \lambda_3^2 u \big) + f_4 u \big( 2f_6 \lambda_1^2 u + \lambda_3 \big( (2 + 2i\sqrt{3}) f_3 \lambda_1 u + i(\sqrt{3} + i) f_1 \lambda_3 u_1 \big) \big) \big) \bigg],
\end{aligned}
\end{equation}
This matrix can be diagonalized by using $U^{\prime}_{\nu 1}$ with the relation
\begin{equation}
\label{e5}
M^{\prime}_\nu= U^{\prime \ *}_{\nu 1}m_{\text{diag}}{U^{\prime\ \dag}_{\nu 1}},
\end{equation}
with
\begin{equation}  
\begin{aligned}
U^{\prime}_{\nu 1}=\left(
\begin{array}{ccc}
 a+bi & c+di & e+fi \\
 g+hi & i^{\prime}+ji & k+li \\
 m+ni & o+pi & q+ri \\
\end{array}
\right).
\end{aligned}
\end{equation}
The mixing matrix is given as
\begin{equation}
\label{eq:paaaaa}
\begin{aligned}
U^{\prime}_{\text{PMNS}} = {U^{'}}^{\dagger}U_{\nu 1}^{\prime} = \begin{pmatrix}
U^{\prime}_{11} & U^{\prime}_{12} & U^{\prime}_{13} \\
U^{\prime}_{21} & U^{\prime}_{22} & U^{\prime}_{23} \\
U^{\prime}_{31} & U^{\prime}_{32} & U^{\prime}_{33} \\
\end{pmatrix}.
\end{aligned}
\end{equation}
where $m_{\text{diag}}=\text{diag}(m_{1},m_{2},m_{3})$ represents the mass matrix that has been diagonalized for the three light neutrinos. The neutrino mass eigenvalues are thus expressed as:
 \begin{align}
 \label{em}
m_{\text{diag}}=\begin{cases}
\text{diag}(m_{1},\sqrt{m_{1}^{2}+\Delta m_{21}^{2}},\sqrt{m_{1}^{2}+\Delta m_{31}^{2}})~~ &\text{in normal hierarchy (NH)} \\
\text{diag}( \sqrt{m_{3}^{2}+\Delta m_{31}^{2}}, \sqrt{m_{3}^{2}+\Delta m_{31}^{2}+\Delta m_{21}^{2}},m_{3})~~ &\text{in inverted hierarchy (IH)}
\end{cases}
 \end{align}
\section{Numerical Analysis}
\label{sec:Numericsal Analysis}
Taking into account the most recent experimental data in NuFIT 6.0 \cite{esteban2025nufit}, the neutrino mixing angles (the solar mixing angle, $\theta_{12}$, the atmospheric mixing angle, $\theta_{23}$, The reactor angle, $\theta_{13}$), the squared mass differences (the solar squared mass difference, $\Delta m_{\text{sol}}^2$ and the atmospheric squared mass difference, $\Delta m_{\text{atm}}^2$) have been precisely measured as shown in table \ref{parameters} and the sum of neutrino masses, $\Sigma m$, is tightly constrained by a combination of cosmological observations and direct detection experiments, with a lower limit of $0.06$ eV and an upper limit of $0.12$ eV \cite{katrin2022direct}. Utilizing equations \ref{eq:e2}, \ref{e5}, \ref{eq:paaaaa} and \ref{eq:e19}, the objective or fitness function $(\eta^{\prime})$ corresponding to these experimental constraints can be expressed as follows.
\begin{equation}
\label{eq:sum}
\begin{aligned}
\eta^{\prime}=\eta^{\prime}_a+\eta^{\prime}_b+\eta^{\prime}_c+\eta^{\prime}_d+\eta^{\prime}_e+\eta^{\prime}_f+\eta^{\prime}_g+\eta^{\prime}_h+\eta^{\prime}_i+\eta^{\prime}_j+\eta^{\prime}_k+\eta^{\prime}_l+\eta^{\prime}_m+\eta^{\prime}_n+\eta^{\prime}_o,
\end{aligned}
\end{equation}
with,
\begin{align*}
\eta^{\prime}_{a}=\bigg[\ &\bigg|\frac{1}{2 \sigma_1}\bigg(\frac{2 \sqrt{3} \sigma_2 \left(\eta_3+1\right) (a+bi)}{\eta_3+2}-\frac{\left(\sqrt{3}-3 i\right) \sigma_2 (g+hi)}{\eta_3+2}\\&+2 \left(\sqrt{3}+3 i\right) \left(\eta_3+1\right) (m+ ni)\bigg)\bigg|-\big|c _{12} c_{13}\big|\ \bigg]^2,
\\
\nonumber
\eta_{b}^{\prime}=\bigg[\ &\bigg|\frac{1}{2 \sigma_1}\bigg(\frac{2 \sqrt{3} \sigma_2 \left(\eta_3+1\right) (c+di)}{\eta_3+2}-\frac{\left(\sqrt{3}-3 i\right) \sigma_2 (i+ji)}{\eta_3+2}\\&+2 \left(\sqrt{3}+3 i\right) \left(\eta_3+1\right) (o+ pi)\bigg)\bigg|-\big|e^{i \beta _1} s_{12} c_{13}\big|\ \bigg]^2,
\\
\nonumber
\eta^{\prime}_{c}=\bigg[\ &\bigg|\frac{1}{2 \sigma_1}\bigg(
\frac{2 \sqrt{3} \sigma_2 \left(\eta_3+1\right) (e+fi)}{\eta_3+2}-\frac{\left(\sqrt{3}-3 i\right) \sigma_2 (k+li)}{\eta_3+2}\\&+2 \left(\sqrt{3}+3 i\right) \left(\eta_3+1\right) (q+ri)
\bigg)\bigg|-\big|e^{i \left(\beta _2+\delta \right)} s_{13}\big|\ \bigg]^2,
\\
\nonumber
\eta^{\prime}_{d}=\bigg[\ &\bigg|\frac{\sqrt{3}}{\sigma_1}\left(\frac{\left(\eta_3+1\right) \sigma_2 (a+bi)}{\eta_3+2}+\frac{\sigma_2 (g+hi)}{\eta_3+2}-2 \left(\eta_3+1\right) (m+ ni)\right)\bigg| \\&-\big|-s_{12} c_{23}-e^{i \delta } s_{13}s_{23}c_{12}\big|\ \bigg]^2,
\\
\nonumber
\eta^{\prime}_{e} = \bigg[\ &\bigg|\frac{\sqrt{3}}{\sigma_1} \left(\frac{(\eta_3 + 1)\, \sigma_2 (c + di)}{\eta_3 + 2} + \frac{\sigma_2 (i^{\prime} + ji)}{\eta_3 + 2} - 2(\eta_3 + 1)(o + pi)\right)\bigg| \\
& - \big| e^{i \beta_1} (c_{12} c_{23} - e^{i \delta} s_{12} s_{13} s_{23}) \big| \ \bigg]^2,
\\
\nonumber
\eta^{\prime}_{f} = \bigg[\ &\bigg|\frac{\sqrt{3}}{\sigma_1} \left(\frac{(\eta_3 + 1)\sigma_2 (e +  fi)}{\eta_3 + 2} + \frac{\sigma_2 (k +  li)}{\eta_3 + 2} - 2 (\eta_3 + 1)(q + ri)\right)\bigg| - \big|e^{i \beta_2} s_{23}c_{13}\big|\ \bigg]^2,
\\
\nonumber
\eta^{\prime}_{g} = \bigg[\ &\bigg| \frac{1}{2 \sigma_1} \bigg( \frac{2 \sqrt{3} \sigma_2 (\eta_3 + 1)(a + bi)}{\eta_3 + 2} - \frac{(\sqrt{3} + 3i) \sigma_2 (g + hi)}{\eta_3 + 2} \\
&+ 2 (\sqrt{3} - 3i)(\eta_3 + 1)(m + ni) \bigg) \bigg|  - \big| s_{12} s_{23} - e^{i \delta} s_{13} c_{12} c_{23} \big| \ \bigg]^2,
\\
\nonumber
\eta^{\prime}_{h} = \bigg[\ &\bigg| \frac{1}{2 \sigma_1} \bigg( \frac{2 \sqrt{3} \sigma_2 (\eta_3 + 1)(c + di)}{\eta_3 + 2} - \frac{(\sqrt{3} + 3i) \sigma_2 (i^{\prime} + ji)}{\eta_3 + 2} \\
& + 2 (\sqrt{3} - 3i)(\eta_3 + 1)(o + pi) \bigg) \bigg| - \big| - e^{i \beta_1} \left( s_{23} c_{12} + e^{i \delta} s_{12} s_{13} c_{23} \right) \big| \ \bigg]^2,
\\
\nonumber
\eta^{\prime}_{i} = \bigg[\ &\bigg| \frac{1}{2 \sigma_1} \bigg( \frac{2 \sqrt{3} \sigma_2 (\eta_3 + 1)(e + fi)}{\eta_3 + 2} - \frac{(\sqrt{3} + 3i) \sigma_2 (k + li)}{\eta_3 + 2} \\
&+ 2 (\sqrt{3} - 3i)(\eta_3 + 1)(q + ri) \bigg) \bigg|  - \big| e^{i \beta_2} c_{13} c_{23} \big| \ \bigg]^2,
\\
\nonumber
\eta^{\prime}_{j}=\bigg[\ &\bigg|\frac{P}{V}\bigg|-\big|m_1 (a- bi)^2+m_2 (c-di)^2+m_3 (e- fi)^2\big|\ \bigg]^2,
\\
\nonumber
\eta^{\prime}_{k}=\bigg[\ &\bigg|\frac{Q}{V}\bigg|-\big|m_1 (a- bi) (g-hi)+m_2 (c-di) (i^{\prime}-ji)+m_3 (e-fi) (k- li)\big|\ \bigg]^2,
\\
\nonumber
\eta^{\prime}_{l}=\bigg[\ &\bigg|\frac{R}{V}\bigg|-\big|m_1 (a- bi) (m-ni)+m_2 (c-di) (o-pi)+m_3 (e-fi) (q-ri)\big|\ \bigg]^2,
\\
\nonumber
\eta^{\prime}_{m}=\bigg[\ &\bigg|\frac{S}{V}\bigg|-\big|m_1 (g-hi)^2+m_2 (i^{\prime}-ji)^2+m_3 (k- li)^2\big|\ \bigg]^2,
\\
\nonumber
\eta^{\prime}_{n}=\bigg[\ &\bigg|\frac{T}{V}\bigg|-\big|m_1 (g- hi) (m-ni)+m_2 (i^{\prime}-ji) (o-pi)+m_3 (k-li) (q-ri)\big|\ \bigg]^2,
\\
\nonumber
\eta^{\prime}_{o}=\bigg[\ &\bigg|\frac{U}{V}\bigg|-\big|m_1 (m-ni)^2+m_2 (o-pi)^2+m_3 (q-ri)^2\big|\ \bigg]^2,
\end{align*}
with $\sigma_1=\left(\eta_3 \left(\eta +\eta_3-6\right)-6\right)$ and $\sigma_2=\left(\eta_3 \left(\eta +\eta_3-4\right)-4\right)$.\\
\begin{table}[htbp]
\centering
\begin{tabular}{ccc}
\hline 
Parameters &Normal Ordering &Inverted Ordering \\ 
\hline 
$sin^2\theta_{12}$ & $0.275\rightarrow 0.345$&$0.275\rightarrow 0.345$ \\ 
$\theta_{12}/^{\circ}$ & $31.63\rightarrow 35.95$&$31.63\rightarrow 35.95$\\ 
$sin^2\theta_{23}$ & $0.435\rightarrow 0.585$&$0.437\rightarrow 0.597$\\ 
$\theta_{23}/^{\circ}$ &$41.3\rightarrow 49.9$&$41.4\rightarrow 50.6$ \\ 
$sin^2\theta_{13}$ & $0.020230\rightarrow 0.02388$&$0.02053\rightarrow 0.02397$ \\ 
$\theta_{13}/^{\circ}$ &$8.19\rightarrow 8.89$&$8.24\rightarrow 8.91$ \\ 
$\delta_{CP}/^{\circ}$ &$124\rightarrow 364$&$201\rightarrow 348 $ \\ 
$\frac{\Delta m^2_{21}}{10^{-5}eV^2}$ & $6.92\rightarrow 8.05$&$6.92\rightarrow 8.05$ \\ 
$\frac{\Delta m^2_{3l}}{10^{-3}eV^2}$ & $+2.451\rightarrow -2.584$&$0.275\rightarrow -2.438$\\
\hline
\end{tabular}
\caption{The table summarizes the best-fit values in $3\sigma$ range for neutrino oscillation parameters: mixing angles ($\sin^2 \theta_{12}$, $\sin^2 \theta_{23}$, $\sin^2 \theta_{13}$), mass-squared differences ($\Delta m^2_{21}$ and $\Delta m^2_{3l}$), and the $\delta_{\text{CP}}$ phase.}
\label{parameters}
\end{table}
We use the PSO algorithm to optimize the objective or fitness functions $\eta^{\prime}$ for the optimal values of the parameters. PSO is an algorithm that optimizes objective functions by simulating the behavior of a swarm of points in a search space. Each point represents a potential solution and moves based on its velocity, influenced by its local best position and the global best position.

\begin{figure}
    \centering
    \includegraphics[width=0.6
    \linewidth]{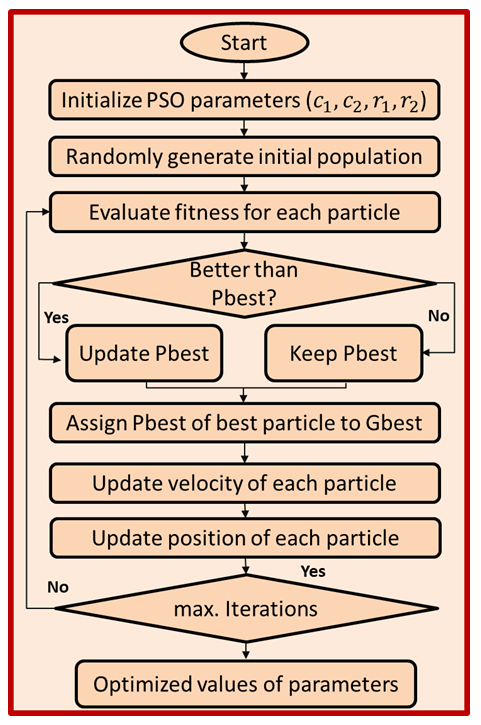}
    \caption{Generic flow chart of PSO}
    \label{fig:PSO1}
\end{figure}

\textbf{Key Steps:}
\begin{enumerate}
\item Initialization: Randomly distribute points in the search space with initial velocities.
\item Fitness Evaluation: Calculate the objective function value for each point.
\item Update Best Positions: Update local best (Pbest) and global best (Gbest) positions based on fitness values.
\item Velocity and Position Update: Adjust velocity and position using equations:
\begin{equation}
\label{eq:PSO1}
\begin{aligned}
\text{Velocity Update:}\  v^{x}_i=wv^{t-1}_i+c_1 r_1(P^{x-1}_{LB}-X^{x-1}_{i})+c_2 r_2(P^{x-1}_{GB}-X^{x-1}_{i}),
\end{aligned}
\end{equation}
\begin{equation}
\label{eq:PSO2}
\begin{aligned}
\text{Position Update:}\ X^{x}_i=X^{x-1}_i+v^{x-1}_i.
\end{aligned}
\end{equation}
\item Iteration: Repeat steps 2-4 until convergence or a predetermined number of iterations.
\end{enumerate}
\textbf{How it Works:}
\begin{itemize}
\item Points adapt their positions and velocities based on individual experiences and neighboring insights.
\item Velocity is controlled by weights $w$, $c_1$, and $c_2$, and random numbers $r_1$ and $r_2$.
\item Velocity limits ensure points don't move too fast or too slow.
\end{itemize}
This process allows the swarm to explore the search space and converge on optimal solutions. The generic flow chart PSO is given in figure \ref{fig:PSO1}.

The objective function $\eta^{\prime}$ is minimized using the PSO technique to find the best fit values of mixing angles, Dirac and Majorana phases, neutrino masses, and the other model parameters for both mass hierarchies are given in tables \ref{tab:optimal values for Normal hierarchy and upper bound limit of} and \ref{tab:optimal values for Normal hierarchy and lower bound limit of}. The lepton mixing matrices for both mass hierarchy are given as:
\begin{equation}
\label{eq:UPMNSmatrix}
\begin{aligned}
|U^{\prime}_{PMNS}|=\begin{cases} \left(
\begin{array}{ccc}
 0.802345 & 0.576507 & 0.154538 \\
 0.377945 & 0.577446 & 0.723681 \\
 0.461953 & 0.578096 & 0.67261 \\
\end{array}
\right)~~ &\text{in normal hierarchy (NH)} \\
\\
\left(
\begin{array}{ccc}
 0.841054 & 0.519855 & 0.149594 \\
 0.455263 & 0.540338 & 0.707651 \\
 0.292169 & 0.661653 & 0.690545 \\
\end{array}
\right)~~ &\text{in inverted hierarchy (IH)}
\end{cases}
\end{aligned}
\end{equation}
\begin{table}[htbp]
\centering
\begin{tabular}{llll}
\hline
 Parameters& Optimal values & Parameters &Optimal values
 \\
\hline
\hline
$\theta_{12}$ & $35.6983^\circ$ & $\theta_{13}$&  $8.89^\circ$ \\
\hline
$\theta_{23}$ & $47.0947^\circ$&  $\delta$& $251.299^\circ$\\
\hline
$\beta_{1}$ & $-57.2958^\circ$ & $\beta_{2}$& $57.2958^\circ$\\
\hline
$m_{1}$ & $0.019998\ eV$ & $m_{2}$& $0.0217904\ eV$\\
\hline
$m_{3}$ & $0.0539715\ eV$ & $H$& $142.028\ GeV$\\
\hline
$\chi$ & $100\ GeV$ & $\lambda_1$& $ -0.162004$\\
\hline
$\lambda_2$ & $0.667057$& $\lambda_3$& $-0.0175654$\\
\hline
$f_1$ & $-8.08086$ & $f_2$& $10.0000 $\\
\hline
$f_3$ & $10.0000$ & $f_4$& $-5.38063$\\
\hline
$f_5$ & $-6.9312$ & $f_6$& $9.57921$\\
\hline
$u$ & $3.94595\times 10^{13}\ eV $ & $u_1$& $1.0000\times 10^{15}\ eV $\\
\hline
$\Lambda$ & $2.4300\times 10^{15}\ eV$ & $\eta_3$& $0.326616$\\
\hline
\end{tabular}
\caption{The optimal values of parameters through PSO for normal mass hierarchy. Where the objective function is minimized upto $3.7196 \times 10^{-16}$.
\label{tab:optimal values for Normal hierarchy and upper bound limit of}}
\end{table}
\begin{table}[htbp]
\centering
\begin{tabular}{llll}
\hline
 Parameters& Optimal values & Parameters &Optimal values
 \\
\hline
\hline
$\theta_{12}$ & $31.7201^\circ$ & $\theta_{13}$&  $8.60336^\circ$ \\
\hline
$\theta_{23}$ & $45.7009^\circ$&  $\delta$& $343.516^\circ$\\
\hline
$\beta_{1}$ & $-57.2958^\circ$ & $\beta_{2}$& $-57.2958^\circ$\\
\hline
$m_{1}$ & $0.0510882\ eV$ & $m_{2}$& $0.051816\ eV$\\
\hline
$m_{3}$ & $0.01\ eV$ & $H$& $142.028\ GeV$\\
\hline
$\chi$ & $100\ GeV$ & $\lambda_1$& $ 0.120072$\\
\hline
$\lambda_2$ & $0.180526$& $\lambda_3$& $0.106161$\\
\hline
$f_1$ & $10$ & $f_2$& $-9.95539 $\\
\hline
$f_3$ & $9.99855$ & $f_4$& $-7.22286$\\
\hline
$f_5$ & $-10$ & $f_6$& $9.35234$\\
\hline
$u$ & $5.69987\times 10^{14}\ eV $ & $u_1$& $8.52501678\times 10^{14}\ eV$\\
\hline
$\Lambda$ & $2.4300\times 10^{15}\ eV$ & $\eta_3$& $0.955218$\\
\hline
\end{tabular}
\caption{The optimal values of parameters through PSO for inverted mass hierarchy. Where the objective function is minimized upto $6.6897 \times 10^{- 14}$.
\label{tab:optimal values for Normal hierarchy and lower bound limit of}}
\end{table}
The Jarlskog invariant, abbreviated as $J_{CP}$, measures how much $CP$ violation there is in neutrino oscillations. It acts as a measure to assess the degree of $CP$ violation in these phenomena is given as \cite{barger1998bi}
\begin{equation}
\label{eq:e73}
\begin{aligned}
J_{\text{CP}}=\text{Im}(U_{12}U_{13}^{*}U_{22}^{*}U_{23})=\begin{cases} -0.027574
~~ &\text{in normal hierarchy (NH)} \\
0.00759102~~ &\text{in inverted hierarchy (IH)}
\end{cases}
\end{aligned}
\end{equation}

\begin{figure}[htbp]
\centering
\includegraphics[width=.8\textwidth]{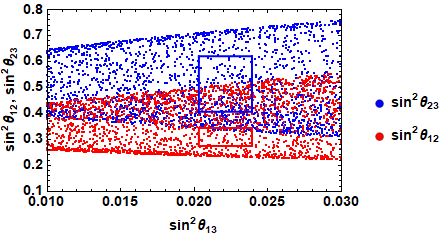}
\caption{Correlation plot illustrating the relationship of $\sin^2 \theta_{13}$ with the atmospheric mixing angle $\sin^2 \theta_{23}$ and the solar mixing angle $\sin^2 \theta_{12}$. This plot provides insight into how variations in $\sin^2 \theta_{13}$ influence the values of the other two mixing angles, highlighting key parameter dependencies within the neutrino mixing framework. The red and blue boxes represent the experimentally allowed $3\sigma$ regions, providing a visual comparison between the model predictions and current neutrino oscillation data.}
\label{fig:correlationplotesss1}
\end{figure}

\begin{figure}[htbp]
\centering
\includegraphics[width=.6\textwidth]{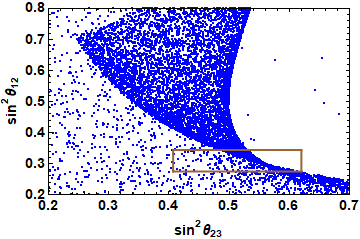}
\caption{The correlation plot between $\sin^2 \theta_{12}$ and $\sin^2 \theta_{23}$ illustrates the interplay between the solar and atmospheric mixing angles. The brown box represents the experimentally allowed $3\sigma$ regions, providing a visual comparison between the model predictions and current neutrino oscillation data.}
\label{fig:correlationplotesss2}
\end{figure}

\begin{figure}[htbp]
\centering
\includegraphics[width=.6\textwidth]{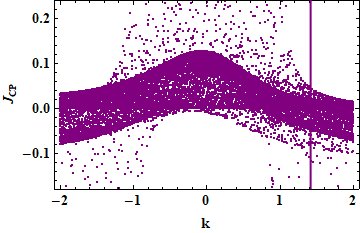}
\caption{The correlation between the Jarlskog invariant and the model parameter $k$. The vertical purple line indicates the specific value of $k$ adopted in our numerical analysis, highlighting its role in generating CP violation consistent with experimental constraints.}
\label{fig:correlationplotesss3}
\end{figure}

\begin{figure}[htbp]
\centering
\includegraphics[width=.6\textwidth]{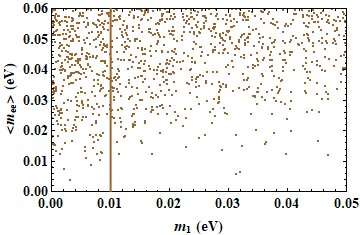}
\caption{Correlation between the effective Majorana mass $\left\langle m_{\text{ee}} \right\rangle$ and the lightest neutrino mass $m_1$. The vertical line indicates the specific value of $m_1$ chosen in our numerical analysis, illustrating its role in shaping the predicted $\left\langle m_{\text{ee}} \right\rangle$ within phenomenologically viable limits.}
\label{fig:correlationplotesss4}
\end{figure}

\begin{figure}[htbp]
\centering
\includegraphics[width=.6\textwidth]{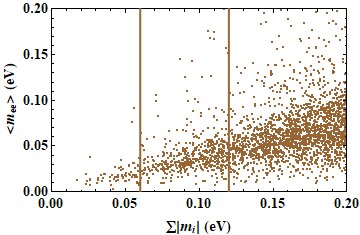}
\caption{Correlation between the effective Majorana mass $\left\langle m_{\text{ee}} \right\rangle$ and the sum of the neutrino masses $\sum |m_i|$. The vertical lines represent the experimentally allowed bounds on $\sum |m_i|$, as reported by cosmological and neutrino oscillation data, providing a comparative reference for the model's predictions.}
\label{fig:correlationplotesss5}
\end{figure}

Moreover, when we additionally introduce an $S_2$ symmetry, which is the permutation group of two elements. In the 3-dimensional representation, this group is characterized by the following generators:
\[
e = \begin{pmatrix} 
1 & 0 & 0 \\ 
0 & 1 & 0 \\ 
0 & 0 & 1 
\end{pmatrix}, \quad
a = \begin{pmatrix} 
1 & 0 & 0 \\ 
0 & 0 & 1 \\ 
0 & 1 & 0 
\end{pmatrix}.
\] 
The imposition of the $S_2$ flavor symmetry leads to the constraints: $f_2 \leftrightarrow f_3$, $f_4 \leftrightarrow f_6$, and $\lambda_2 \leftrightarrow \lambda_3$. Consequently, the neutrino mass matrix is expressed as follows:
\begin{equation}
\label{eq:e20}
\begin{aligned}
M_\nu=
  \begin{pmatrix}
    A & B & B\\
       B& C & D\\
      B & D & C
  \end{pmatrix},
\end{aligned}
\end{equation}
where,
\begin{equation}
\label{eq:e21}
\begin{aligned}
A=\frac{H^2 \chi ^2 \left(2 f_1 \lambda^2_{2} u_1-4 f_3\lambda_1\lambda_2 u+\lambda^2_{1} (f_4 u+f_5 u_1)\right)}{\Lambda^2 \left(2f_3^2 u^2-f_1 u_1 (f_4 u+f_5 u_1)\right)},
\end{aligned}
\end{equation}
\begin{equation}
\label{eq:e22}
\begin{aligned}
B=\frac{H^2 \chi^2 \left(-\lambda_2(f_1 \lambda_2 u_1+f_3 \lambda_1 u)+f_4\lambda_1^2 u+f_5 \lambda_1^2 u_1\right)}{\Lambda^2 \left(2 f_3^2 u^2-f_1 u_1 (f_4 u+f_5 u_1)\right)},
\end{aligned}
\end{equation}
\begin{equation}
\label{eq:e23}
\begin{aligned}
C=\frac{H^2 \chi ^2 \left(\lambda_2^2 \left(f_1 u_1 (f_4 u+2f_5 u_1)-3 f_3^2 u^2\right)+2 f_3\lambda_1 \lambda_2 u (f_5 u_1-f_4 u)+\lambda_1^2 \left(f_5^2 u_1^2-f_4^2 u^2\right)\right)}{\Lambda ^2 (f_4 u-f_5 u_1) \left(f_1 u_1 (f_4 u+f_5 u_1)-2f_3^2 u^2\right)},
\end{aligned}
\end{equation}
\begin{equation}
\label{eq:e24}
\begin{aligned}
D=&-\frac{H^2 \chi^2 (\lambda_2^2(f_1 u_1 (2 f_4 u+f_5 u_1)-3f_3^2 u^2)}{\Lambda^2 (f_4 u-f_5 u_1)(f_1 u_1 (f_4 u+f_5 u_1)-2f_3^2 u^2)}\\ &
-\frac{H^2 \chi^2(2 f_3\lambda_1\lambda_2 u (f_4 u-f_5 u_1)+\lambda_1^2 (f_4 u+f_5 u_1) (f_4 u-f_5 u_1))}{\Lambda^2 (f_4 u-f_5 u_1)(f_1 u_1 (f_4 u+f_5 u_1)-2f_3^2 u^2)}.
\end{aligned}
\end{equation}
The associated diagonalized matrix the $M_\nu$ is given as
\begin{equation}
\label{eq:e25}
\begin{aligned}
U_{\nu} = \begin{pmatrix}
\frac{q_{1}}{\sqrt{2 + q_{1}^{2}}} & 0 & - \frac{q_{2}}{\sqrt{2 + q_{2}^{2}}} \\
\frac{1}{\sqrt{2 + q_{1}^{2}}} & - \frac{1}{\sqrt{2}} & \frac{1}{\sqrt{2 + q_{2}^{2}}} \\
\frac{1}{\sqrt{2 + q_{1}^{2}}} & \frac{1}{\sqrt{2}} & \frac{1}{\sqrt{2 + q_{2}^{2}}} \\
\end{pmatrix},
\end{aligned}
\end{equation}
with
\begin{equation}
\label{eq:e26}
\begin{aligned}
q_{1} = \frac{\left( A - C - D + \sqrt{A^{2} + 8B^{2} - 2A(C + D) + (C + D)^{2}} \right)}{2B},
\end{aligned}
\end{equation}
\begin{equation}
\label{eq:e27}
\begin{aligned}
q_{2} = \frac{\left( - A + C + D + \sqrt{A^{2} + 8B^{2} - 2A(C + D) + (C + D)^{2}} \right)\ }{2B}.
\end{aligned}
\end{equation}
The mass eigenvalues are
\begin{equation}
\label{eq:e28}
\begin{aligned}
m_{i} = {U^{\dagger}}_{\nu}M_{\nu}U_{\nu},
\end{aligned}
\end{equation}
\begin{equation}
\label{eq:e29}
\begin{aligned}
m_{1,3} = \frac{1}{2}\left( A + C + D \pm \sqrt{8B^{2} + ( - A + C + D)^{2}} \right),\quad m_{2} = C - D.
\end{aligned}
\end{equation}
Considering a unitary matrix 
\begin{equation}
\label{eq:e30}
\begin{aligned}
U_{Z} = \begin{pmatrix}
a & b & c \\
d & e & f \\
g & h & i \\
\end{pmatrix},
\end{aligned}
\end{equation}
where,
\begin{equation}
\label{eq:e31}
\begin{aligned}
a=0,
\end{aligned}
\end{equation}
\begin{equation}
\label{eq:e32}
\begin{aligned}
b = - \frac{3i\sqrt{2 + q_{1}^{2}} + \sqrt{3}\sqrt{2 + q_{1}^{2}} + 3q_{2}\sqrt{2 + q_{2}^{2}} - i\sqrt{3}q_{2}\sqrt{2 + q_{2}^{2}}}{2\sqrt{3}\left( q_{1} + q_{2} \right)\sqrt{1 + q_{2}^{2}}},
\end{aligned}
\end{equation}
\begin{equation}
\label{eq:e33}
\begin{aligned}
c = - \frac{- 3i\sqrt{2 + q_{1}^{2}}q_{2} + \sqrt{3}\sqrt{2 + q_{1}^{2}}q_{2} + 3\sqrt{2 + q_{2}^{2}} + i\sqrt{3}\sqrt{2 + q_{2}^{2}}}{2\sqrt{3}\left( q_{1} + q_{2} \right)\sqrt{1 + q_{2}^{2}}},
\end{aligned}
\end{equation}
\begin{equation}
    \label{eq:e34}
    \begin{aligned}
        d = - \frac{1}{\sqrt{2}},
    \end{aligned}
\end{equation}
\begin{equation}
    \label{eq:e35}
    \begin{aligned}
e = - \frac{\frac{iq_1\left( 3iq_{2} + \sqrt{3}q_{2} \right)}{2\sqrt{3}\sqrt{2 + q_{1}^{2}}\sqrt{1 + q_{2}^{2}}} + \frac{\left( 3i + \sqrt{3} \right)q_{2}}{2\sqrt{3}\sqrt{1 + q_{2}^{2}}\sqrt{2 + q_{2}^{2}}}}{2\left( \frac{q_{1}}{\sqrt{2 + q_{1}^{2}}\sqrt{2 + q_{2}^{2}}} + \frac{q_{2}}{\sqrt{2 + q_{1}^{2}}\sqrt{2 + q_{2}^{2}}} \right)},
    \end{aligned}
\end{equation}
\begin{equation}
    \label{eq:e36}
    \begin{aligned}
f = - \frac{- \frac{i\left( - 3i + \sqrt{3} \right)q_{1}}{2\sqrt{3}\sqrt{2 + q_{1}^{2}}\sqrt{1 + q_{2}^{2}}} + \frac{q_{2}\left( - 3iq_{2} + \sqrt{3}q_{2} \right)}{2\sqrt{3}\sqrt{1 + q_{2}^{2}}\sqrt{2 + q_{2}^{2}}}}{2\left( \frac{q_{1}}{\sqrt{2 + q_{1}^{2}}\sqrt{2 + q_{2}^{2}}} + \frac{q_{2}}{\sqrt{2 + q_{1}^{2}}\sqrt{2 + q_{2}^{2}}} \right)},        
    \end{aligned}
\end{equation}
\begin{equation}
    \label{eq:e37}
    \begin{aligned}
g = \frac{1}{\sqrt{2}},        
    \end{aligned}
\end{equation}
\begin{equation}
    \label{eq:e38}
    \begin{aligned}
h = - \frac{\frac{iq_{1}\left( 3iq_{2} + \sqrt{3}q_{2} \right)}{2\sqrt{3}\sqrt{2 + q_{1}^{2}}\sqrt{1 + q_{2}^{2}}} + \frac{\left( 3i + \sqrt{3} \right)q_{2}}{2\sqrt{3}\sqrt{1 + q_{2}^{2}}\sqrt{2 + q_{2}^{2}}}}{2\left( \frac{q_{1}}{\sqrt{2 + q_{1}^{2}}\sqrt{2 + q_{2}^{2}}} + \frac{q_2}{\sqrt{2 + q_{1}^{2}}\sqrt{2 + q_{2}^{2}}} \right)},        
    \end{aligned}
\end{equation}
\begin{equation}
    \label{eq:e39}
    \begin{aligned}
i = - \frac{- \frac{i( - 3i + \sqrt{3})q_{1}}{2\sqrt{3}\sqrt{2 + q_{1}^{2}}\sqrt{1 + q_{2}^{2}}} + \frac{q_2( - 3iq_{2} + \sqrt{3}q_{2})}{2\sqrt{3}\sqrt{1+q_{2}^{2}}\sqrt{2 + q_{2}^{2}}}}{2\left( \frac{q_{1}}{\sqrt{2 + q_{1}^{2}}\sqrt{2 + q_{2}^{2}}} + \frac{q_{2}}{\sqrt{2 + q_{1}^{2}}\sqrt{2 + q_{2}^{2}}} \right)},        
    \end{aligned}
\end{equation}
with $k\equiv q_2=2/q_1$ which is obtained by the unitary condition of matrix $U_z$ and also from equations \ref{eq:e26} and \ref{eq:e27}.
The associated neutrino mixing matrix is given as 
\begin{equation}
    \label{eq:e40}
    \begin{aligned}
U_{\nu}^{'} = \ {U_{Z}}^{\dagger}.\text{U}_{\nu},
    \end{aligned}
\end{equation}
\begin{equation}
    \label{eq:e41}
    \begin{aligned}
U_{\nu}^{'} = \begin{pmatrix}
0 & 1 & 0 \\
 - \frac{3i + \sqrt{3}}{2\sqrt{3}\sqrt{1 + k^{2}}} & 0 & - \frac{i(3ik + \sqrt{3}k)}{2\sqrt{3}\sqrt{1 + k^{2}}} \\
 - \frac{- 3ik + \sqrt{3}k}{2\sqrt{3}\sqrt{1 + k^{2}}} & 0 & \frac{i( - 3i + \sqrt{3})}{2\sqrt{3}\sqrt{1 + k^{2}}} \\
\end{pmatrix}.
    \end{aligned}
\end{equation}
The matrices $U_L$, $U_R$ which are used to diagonalize the charged lepton mass matrix are given as
\begin{equation}
    \label{eq:e42}
    \begin{aligned}
U_{L} = \begin{pmatrix}
1 & P & Q \\
Q & 1 & P \\
P & Q & 1 \\
\end{pmatrix}, \qquad U_{R} =1,
    \end{aligned}
\end{equation}
where,
\begin{equation}
    \label{eq:e43}
    \begin{aligned}
P = \frac{\left( 1 + i\sqrt{3} \right)\eta_3 + \eta_2\left( 1 - i\sqrt{3} + 2\eta_3 \right)}{6 + 4\eta_3 + 2\eta_2\left( 2 + \eta_3 \right)},
    \end{aligned}
\end{equation}
\begin{equation}
    \label{eq:e44}
    \begin{aligned}
Q = \frac{\left( 1 - i\sqrt{3} \right)\eta_3 + \eta_2\left( 1 + i\sqrt{3} + 2\eta_3 \right)}{6 + 4\eta_3 + 2\eta_2\left( 2 + \eta_3 \right)},
    \end{aligned}
\end{equation}
by unitary condition $U_L$, one can get the relation given in equation \ref{eq:e45} then $U_L$ matrix will become like as equation \ref{eq:e47},
\begin{equation}
    \label{eq:e45}
    \begin{aligned}
\eta_2 = \frac{2\eta_3 - \eta_3^{2} - \eta_3\eta}{2\left( 2 + \eta_3 \right)},
    \end{aligned}
\end{equation}
with,
\begin{equation}
    \label{eq:e46}
    \begin{aligned}
\eta = \sqrt{- 12 + \left( - 12 + \eta_3 \right)\eta_3},
    \end{aligned}
\end{equation}
\begin{equation}
    \label{eq:e47}
    \begin{aligned}
U_{L} = \begin{pmatrix}
1 & U_{13}^{l} & U_{12}^{l} \\
U_{12}^{l} & 1 & U_{13}^{l} \\
U_{13}^{l} & U_{12}^{l} & 1 \\
\end{pmatrix},
    \end{aligned}
\end{equation}
where,
\begin{equation}
    \label{eq:e48}
    \begin{aligned}
U_{13}^{l} = \frac{\eta_3\left( - 6 - 2i\sqrt{3} + \eta - i\sqrt{3}\eta + \eta_3\left( - 5 - 3i\sqrt{3} + 2\eta_3 + 2\eta \right) \right)}{2\left( 2 + \eta_3 \right)\left( - 6 + \eta_3\left( - 6 + \eta_3 + \eta \right) \right)},
    \end{aligned}
\end{equation}
\begin{equation}
 \label{eq:e49}
 \begin{aligned}
U_{12}^{l} = \frac{\eta_3\left( - 6 + 2i\sqrt{3} + \eta + i\sqrt{3}\eta + \eta_3\left( - 5 + 3i\sqrt{3} + 2\eta_3 + 2\eta \right) \right)}{2\left( 2 + \eta_3 \right)\left( - 6 + \eta_3\left( - 6 + \eta_3 + \eta \right) \right)},
\end{aligned}
\end{equation}
and
\begin{equation}
\label{eq:e50}
\begin{aligned}
U^{'} = U_{\text{OL}}^{'}.{U_{L}}^{\dagger},
\end{aligned}
\end{equation}
\begin{equation}
\label{eq:e51}
\begin{aligned}
U^{'} = \begin{pmatrix}
Y_{a} & Y_{b} & Y_{c} \\
Y_{d} & Y_{e} & Y_{f} \\
Y_{g} & Y_{h} & Y_{i} \\
\end{pmatrix},
\end{aligned}
\end{equation}
where,
\begin{equation}
\label{eq:e52}
\begin{aligned}
Y_{a} = \frac{\sqrt{3\left( 1 + \eta_3 \right)\left( - 4 + \eta_3^{2} + \eta_3( - 4 + \eta) \right)}}{\left( 2 + \eta_3 \right)\left( - 6 + \eta_3^{2} + \eta_3( - 6 + \eta) \right)},
\end{aligned}
\end{equation}
\begin{equation}
\label{eq:e53}
\begin{aligned}
Y_{b} = \frac{\sqrt{3\left( 1 + \eta_3 \right)\left( - 4 + \eta_3^{2} + \eta_3( - 4 + \eta) \right)}}{\left( 2 + \eta_3 \right)\left( - 6 + \eta_3^{2} + \eta_3( - 6 + \eta) \right)},
\end{aligned}
\end{equation}
\begin{equation}
\label{eq:e54}
\begin{aligned}
Y_{c} = \frac{\sqrt{3\left( 1 + \eta_3 \right)\left( - 4 + \eta_3^{2} + \eta_3( - 4 + \eta) \right)}}{\left( 2 + \eta_3 \right)\left( - 6 + \eta_3^{2} + \eta_3( - 6 + \eta) \right)},
\end{aligned}
\end{equation}
\begin{equation}
\label{eq:e55}
\begin{aligned}
Y_{d} = - \frac{\left( 3i + \sqrt{3} \right)\left( - 4 + \eta_3^{2} + \eta_3( - 4 + \eta) \right)}{2\left( 2 + \eta_3 \right)\left( - 6 + \eta_3^{2} + \eta_3( - 6 + \eta) \right)},
\end{aligned}
\end{equation}
\begin{equation}
\label{eq:e56}
\begin{aligned}
Y_{e} = \frac{\sqrt{3\left( - 4 + \eta_3^{2} + \eta_3( - 4 + \eta) \right)}}{\left( 2 + \eta_3 \right)\left( - 6 + \eta_3^{2} + \eta_3( - 6 + \eta) \right)},
\end{aligned}
\end{equation}
\begin{equation}
\label{eq:e57}
\begin{aligned}
Y_{f} = - \frac{\left( - 3i + \sqrt{3} \right)\left( - 4 + \eta_3^{2} + \eta_3( - 4 + \eta) \right)}{2\left( 2 + \eta_3 \right)\left( - 6 + \eta_3^{2} + \eta_3( - 6 + \eta) \right)},
\end{aligned}
\end{equation}
\begin{equation}
\label{eq:e58}
\begin{aligned}
Y_{g} = \frac{\left( - 3i + \sqrt{3} \right)\left( 1 + \eta_3 \right)}{- 6 + \eta_3^{2} + \eta_3( - 6 + \eta)},\qquad Y_{h} = - \frac{2\sqrt{3\left( 1 + \eta_3 \right)}}{- 6 + \eta_3^{2} + \eta_3( - 6 + \eta)},
\end{aligned}
\end{equation}
\begin{equation}
\label{eq:e59}
\begin{aligned}
Y_{i} = \frac{\left( 3i + \sqrt{3} \right)\left( 1 + \eta_3 \right)}{- 6 + \eta_3^{2} + \eta_3( - 6 + \eta)}.
\end{aligned}
\end{equation}
$U_{PMNS}$ matrix is given as
\begin{equation}
\label{eq:e60}
\begin{aligned}
U_{\text{PMNS}} = {U^{'}}^{\dagger}U_{\nu}^{'},
\end{aligned}
\end{equation}
\begin{equation}
\label{eq:e61}
\begin{aligned}
U_{\text{PMNS}} = \begin{pmatrix}
U_{11} & U_{12} & U_{13} \\
U_{21} & U_{22} & U_{23} \\
U_{31} & U_{32} & U_{33} \\
\end{pmatrix},
\end{aligned}
\end{equation}
where,
\begin{equation}
\label{eq:e62}
\begin{aligned}
U_{11} = \frac{\sqrt{3}\left( - 4(1 + k) + \eta_3\left( - 4 - 6k + (1 - 2k)\eta_3 + \eta \right) \right)}{\sqrt{1 + k^{2}}\left( 2 + \eta_3 \right)\left( - 6 + \eta_3\left( - 6 + \eta_3 + \eta \right) \right)},
\end{aligned}
\end{equation}
\begin{equation}
\label{eq:e63}
\begin{aligned}
U_{13} =\frac{i \sqrt{3} \left(\eta_3 \left(k (\eta -4)+(k+2) \eta_3+6\right)-4 k+4\right)}{\sqrt{k^2+1} \left(\eta_3+2\right) \sigma_1},
\end{aligned}
\end{equation}
\begin{equation}
\label{eq:e64}
\begin{aligned}
U_{21} =\frac{2 \left(\sqrt{3}-3 i\right) k \left(\eta_3+1\right) \left(\eta_3+2\right)-3 i \eta  \eta_3-\sqrt{3} \eta  \eta_3-\left(\sqrt{3}+3 i\right) \left(\left(\eta_3-4\right) \eta_3-4\right)}{2 \sqrt{k^2+1} \left(\eta_3+2\right) \sigma_1},
\end{aligned}
\end{equation}
\begin{equation}
\label{eq:e65}
\begin{aligned}
U_{23} =\frac{-i \sqrt{3} k \eta  \eta_3+3 k \eta  \eta_3+\left(3-i \sqrt{3}\right) k \left(\left(\eta_3-4\right) \eta_3-4\right)+\left(-6-2 i \sqrt{3}\right) \left(\eta_3+1\right) \left(\eta_3+2\right)}{2 \sqrt{k^2+1} \left(\eta_3+2\right) \sigma_1},
\end{aligned}
\end{equation}
\begin{equation}
\label{eq:e66}
\begin{aligned}
U_{31} =&(\eta_3 ((2(\sqrt{3}+3 i)k+3i-\sqrt{3}) \eta_3+6 (\sqrt{3}+3 i) k-(\sqrt{3}-3 i) (\eta -4))\\&+4 ((\sqrt{3}+3 i) k-3 i+\sqrt{3}))/(2 \sqrt{k^2+1} (\eta_3+2) (\eta_3 (\eta +\eta_3-6)-6)),
\end{aligned}
\end{equation}
\begin{equation}
\label{eq:e67}
\begin{aligned}
U_{33} =-\frac{i \left(\frac{\left(\sqrt{3}-3 i\right) k \sigma_2}{\eta_3+2}+2 \left(\sqrt{3}+3 i\right) \left(\eta_3+1\right)\right)}{2 \sqrt{k^2+1} \left(\eta_3 \left(\eta +\eta_3-6\right)-6\right)},
\end{aligned}
\end{equation}
\begin{equation}
\label{eq:e68}
\begin{aligned}
U_{12} =U_{22} =U_{32} =\frac{\left(-\eta +\eta_3+12\right) \eta_3+12}{6 \sqrt{3} \left(\eta_3+2\right)},
\end{aligned}
\end{equation}
In neutrino sector, we have total 6 number of free parameters, which are $A$, $B$, $C$, $D$, $k$ and $\eta_3$. In which $k$ and $\eta_3$ are fixed for the precise measurements of mixing matrix $U_{PMNS}$ and values of $A$, $B$, $C$ and $D$ are measured for both mass hierarchies of neutrinos. The correlations plots among the leptonic mixing parameters of our model are shown in figure \ref{fig:correlationplotesss1}, \ref{fig:correlationplotesss2}, \ref{fig:correlationplotesss3}, \ref{fig:correlationplotesss4} and \ref{fig:correlationplotesss5}.


\begin{figure}[htbp]
\centering
\includegraphics[width=0.6\textwidth]{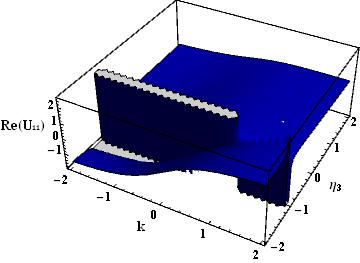}
\caption{Real part of the element $U_{11}$ of the leptonic mixing matrix as a function of the model parameters $k$ and $\eta_3$, where $k \in (-2,2)$ and $\eta_3 \in (-2,2)$. The blue surface illustrates the variation of $\text{Re}(U_{11})$ across the parameter space. The white dot marks the specific parameter values selected for our numerical analysis.}
\label{fig:U1}
\end{figure}

\begin{figure}[htbp]
\centering
\includegraphics[width=0.6\textwidth]{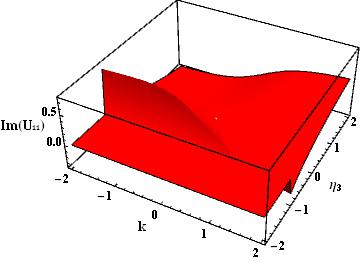}
\caption{Imaginary part of the element $U_{11}$ of the leptonic mixing matrix as a function of $k$ and $\eta_3$, over the same range $k \in (-2,2)$ and $\eta_3 \in (-2,2)$. The red surface highlights the behavior of $\text{Im}(U_{11})$ within the chosen parameter space. As in the previous figure, the white dot indicates the benchmark point used in our analysis.}
\label{fig:U2}
\end{figure}

To determine the precise value $U_{11}=0.805$, accompanied by an experimental observation, a specific selection of the point at which $k=\sqrt{2}$ and $\eta_3=-0.000743889+0.000785038i$ ($\eta_2=-0.00105178 - 0.000251124i$ is estimated from equation \ref{eq:e45}) is required. It is also represented in white dots in the real and imaginary parts of $U_{11}$ in figure \ref{fig:U1} and \ref{fig:U2}.

It should be noted that this model allows for a variety of parameter selections, and the given situation is only one example illustrating the presence of model parameters that 
are consistent with experimental results. The graph of real and imaginary parts of $U_{11}$ in terms of $\eta_3$ and
$k$ is given in figure \ref{fig:U1} and \ref{fig:U2}, where the ranges of $\eta_3$ and k are from $-2$ to $2$.
The $U_{PMNS}$ and its absolute values are given by in equation \ref{eq:e69} and \ref{eq:e70}.
\begin{equation}
\label{eq:e69}
\begin{aligned}
U_{PMNS}=\left(
\begin{array}{ccc}
 0.805\, & 0.577&  0.137988 i \\
 -0.402851+0.119716 i & 0.577 & 0.696899\, -0.0691182 i \\
 -0.402149-0.119716 i & 0.577 & -0.697328-0.0688701 i \\
\end{array}
\right)
\end{aligned},
\end{equation}
\begin{equation}
\label{eq:e70}
\begin{aligned}
|U_{PMNS}|=\left(
\begin{array}{ccc}
 0.805 & 0.577 & 0.137989 \\
 0.420263 & 0.577 & 0.70031 \\
 0.41959 & 0.577 & 0.70072 \\
\end{array}
\right)
\end{aligned},
\end{equation}
Using equation \ref{eq:e2} and following the relations which are given in \ref{eq:e71}, we get following mixing angles given in \ref{eq:e72}
\begin{equation}
\label{eq:e71}
\begin{aligned}
\text{sin}\theta_{13} = \left| U_{13} \right|,\quad \text{sin}\theta_{23} = \frac{\left| U_{23} \right|}{\sqrt{1 - \left| U_{13} \right|^{2}}},\quad \text{sin}\theta_{12} = \frac{\left| U_{12} \right|}{\sqrt{1 - \left| U_{13} \right|^{2}}}
\end{aligned}
\end{equation}
\begin{equation}
\label{eq:e72}
\begin{aligned}
\theta_{13} = {7.9315}^{{^\circ}},\quad \theta_{23} = {44.9980}^{{^\circ}},\quad \theta_{12} = {35.6320}^{{^\circ}}
\end{aligned}
\end{equation}
These results align closely with the recent data on neutrino 
mixing angles. In addition, when we compare the lepton mixing 
matrix presented in equation \ref{eq:e69} with the conventional 
parameterization in equation \ref{eq:e2}, we observe that the Majorana 
phases, specifically $\beta_1$ and $\beta_2$, become zero and the leptonic Dirac $CP$-violating phase $\delta$ takes on the value of $-\pi/2$, and the Jarlskog invariant is measured to be $J_{\text{CP}}=-0.032$. The Majorana phases $\beta_1$ and $\beta_2$ are eliminated after imposing the $S_2$ symmetry and the appropriate choice of the model parameters $k$ and $\eta_3$. Prior to the implementation of this symmetry, the phases are generally non-zero. However, the symmetry constraints enforce a particular relationship among the parameters, leading to the cancellation of these phases. The elimination is not coincidental, but rather a consequence of the underlying symmetry structure and parameter selection in the model. The disappearance of Majorana phases has an immediate consequence for the effective Majorana mass $\langle m_{ee} \rangle$, which is an active player for neutrinoless double beta decay and is sensitive to CP-violating effects from these phases. Quantitatively, we find that the effective neutrino mass parameter is reduced from $\langle m_{ee} \rangle = 21.4051~\text{meV}$ for the normal hierarchy (NH) and $\langle m_{ee} \rangle = 50.3654~\text{meV}$ for the inverted hierarchy (IH), to $\langle m_{ee} \rangle = 13.1448~\text{meV}$ (NH) and $12.6608~\text{meV}$ (IH), respectively, due to these vanishing phases (see section \ref{sec:Effec}). This reproduces significant phenomenological implications of imposing such a symmetry and emphasizes the model's predictive power.

By using $k\equiv q_2=\sqrt{2}$  in equation \ref{eq:e26} and \ref{eq:e27}, one can get, following mass terms for neutrinos
\begin{equation}
\label{eq:e76}
\begin{aligned}
m_{1} = A + \sqrt{2}B,\quad m_{2} = - A + 2C,\quad m_{3} = A - \sqrt{2}B
\end{aligned}
\end{equation}

By using equations \ref{eq:e76} and incorporating the experimental values of the neutrino mass squared differences within the normal hierarchy, as reported in reference \cite{esteban2024nufit} (specifically, $\Delta m_{21}^2=7.49\times10^{-5}$ $eV^{2}$, $\Delta m_{32}^2=2.513\times10^{-3}$ $eV^{2}$, then one can get following values of  B and C in terms of A
\begin{equation}
\label{eq:e77}
\begin{aligned}
B=-\frac{4.4424\times 10^{-4}\ eV^2}{A}, \quad C=\frac{1}{2}\bigg(A+\sqrt{A^2+\frac{3.9470\times 10^{-7}}{A^2}-1.1816\times 10^{-3}}\bigg)
\end{aligned}
\end{equation}
Now, using above values of B and C from equations \ref{eq:e77}, the masses given in \ref{eq:e76}
\begin{equation}
\label{eq:e79}
\begin{aligned}
m_{1,3}=A\mp \frac{6.2825 \times10^{-4}\ eV^{2}}{A},\quad m_2=\sqrt{A^2+\frac{3.9470\times 10^{-7}}{A^2}-1.1816\times 10^{-3}}
\end{aligned}
\end{equation}

For normal hierarchy, we make the safe assumption that $m_1=10^{-2}$ $eV$ which is sufficiently small in order to get precise values of the model parameters. As a result, the following explicit values are given for other parameters: $A=3.0559 \times 10^{-2}\ eV$, $B=-1.4537\times 10^{-2}\ eV$, $C=2.1892 \times10^{-2}\ eV$, and
the following explicit values are given for the remaining neutrino masses: $m_2=1.3225\times10^{-2}$ $eV$, $m_3=5.1118 \times10^{-2}$ $eV$
and sum of neutrino masses is:
$\sum_{}^{}m = \sum_{i = 1}^{3}\left| m_{i} \right| =7.4342 \times10^{-2}$ $eV$. The optimal values for the model parameters $f_1$, $f_3$, $f_4$, $f_5$, $\lambda_1$, $\lambda_2$, $u_1$ and $u$ are estimated by using PSO technique shown in table \ref{tab:optimal values for Normal parameters}.
\begin{table}[htbp]
\centering
\begin{tabular}{llll}
\hline
 Parameters& Optimal values & Parameters &Optimal values
 \\
\hline
\hline
$f_1$ & $1.6822$ & $f_3$&  $-10.0000$ \\
\hline
$f_4$ & $-10.0000$&  $f_5$& $-9.9782 $\\
\hline
$\lambda_1$ & $5.0607 \times 10^{-29}$ & $\lambda_2$ & $9.5720 \times 10^{-29}$\\
\hline
$u_1$ & $1.0000\times 10^{14}\ eV$ & $u$ & $3.7405 \times 10^{13}\ eV$\\
\hline
$H$ & $142.028\ GeV$ & $\chi$ & $100\ GeV$\\
\hline
$\Lambda$ & $2.4300\times 10^{15}\ eV$ &  & \\
\hline
\end{tabular}
\caption{The optimal values of model's parameters for normal mass hierarchy.
\label{tab:optimal values for Normal parameters}}
\end{table}
By using equations \ref{eq:e76} and incorporating the experimental values of the neutrino mass squared differences within the inverted hierarchy, as reported in reference \cite{esteban2024nufit} (specifically, $\Delta m_{21}^2=7.49\times10^{-5}$ $eV^{2}$, $\Delta m_{32}^2=-2.510\times10^{-3}$ $eV^{2}$, one can get following values of  B and C in terms of A
\begin{equation}
\label{eq:e84}
\begin{aligned}
B=\frac{4.3047 \times 10^{-4} \ eV^2}{A}, \quad C=\frac{1}{2}\bigg(A+\sqrt{A^2+\frac{3.7061\times 10^{-7}}{A^2}+1.2924\times 10^{-3}}\bigg)
\end{aligned}
\end{equation}
Now, using above values of B and C from equations \ref{eq:e84}, the masses are given in \ref{eq:e76} 
\begin{equation}
\label{eq:e86}
\begin{aligned}
m_{1,3}=A\pm\frac{6.0878 \times 10^{-4}\ eV^2}{A}, \quad m_2=&\sqrt{A^2+\frac{3.7061\times 10^{-7}}{A^2}+1.2924\times 10^{-3}}
\end{aligned}
\end{equation}

\begin{figure}[htbp]
\centering
\includegraphics[width=0.7\textwidth]{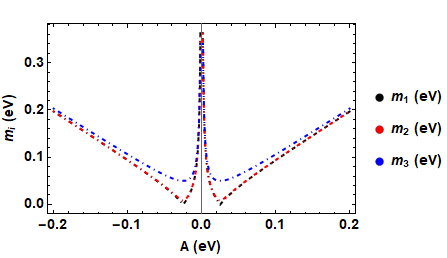}
\caption{Variation of neutrino mass eigenstates $m_1$, $m_2$, and $m_3$ (in eV) as a function of the parameter $A$ in the case of normal mass hierarchy. The range of $A$ is from $-0.2$ to $0.2$. The black, red, and blue dot-dashed curves represent $m_1$, $m_2$, and $m_3$, respectively.}
\label{fig:Norm and invert mass D and sum1}
\end{figure}

\begin{figure}[htbp]
\centering
\includegraphics[width=0.7\textwidth]{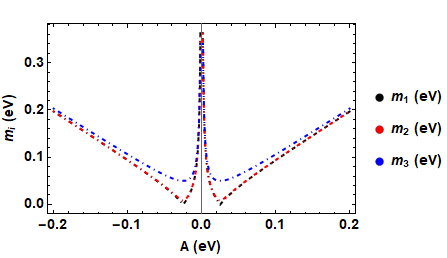}
\caption{Variation of neutrino mass eigenstates $m_1$, $m_2$, and $m_3$ (in eV) as a function of the parameter $A$ in the case of inverted mass hierarchy. As before, $A$ varies from $-0.2$ to $0.2$. The black, red, and blue dot-dashed lines represent $m_1$, $m_2$, and $m_3$, respectively.}
\label{fig:Norm and invert mass D and sum2}
\end{figure}

\begin{figure}[htbp]
\centering
\includegraphics[width=0.6\textwidth]{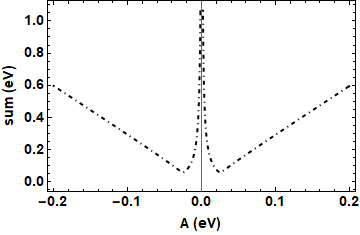}
\caption{The total sum of neutrino masses $\sum m_i$ (in eV) plotted against the model parameter $A$ for the normal hierarchy case.}
\label{fig:Norm and invert mass D and sum3}
\end{figure}

\begin{figure}[htbp]
\centering
\includegraphics[width=0.6\textwidth]{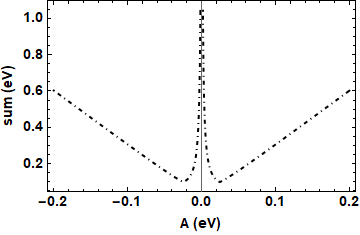}
\caption{The total sum of neutrino masses $\sum m_i$ (in eV) versus the parameter $A$ in the case of inverted mass hierarchy. This figure complements the previous one by showing the behavior under the alternative hierarchy scenario, with the same $A$ range from $-0.2$ to $0.2$.}
\label{fig:Norm and invert mass D and sum4}
\end{figure}

For inverted hierarchy, we also make the safe assumption here, that $m_3=10^{-2}$ $eV$ which is sufficiently small in order to get precise values of the model parameters. As a result, the following explicit values are given for other parameters: $A=3.0175 \times10^{-2}\ eV$, $B=1.4266 \times10^{-2}\ eV$, $C=4.0632\ \times10^{-2} eV$, and the following explicit values are given for the remaining neutrino masses:
$m_1=5.0350\ \times10^{-2}$ $eV$, $m_2=5.1088\times10^{-2}$ $eV$
and sum of neutrino masses is:
$\sum_{}^{}m = \sum_{i = 1}^{3}\left| m_{i} \right| = 1.1144 \times10^{-1}$ $eV$. The variation of masses of neutrinos and $\sum_{}^{}m$ with the value of $A$ are shown in figures \ref{fig:Norm and invert mass D and sum1}, \ref{fig:Norm and invert mass D and sum2}, \ref{fig:Norm and invert mass D and sum3} and \ref{fig:Norm and invert mass D and sum4}.

The optimal values for the model parameters $f_1$, $f_3$, $f_4$, $f_5$, $\lambda_1$, $\lambda_2$, $u_1$ and $u$ are estimated by using the PSO technique are shown in table \ref{tab:optimal values for inverted parameters}.
\begin{table}[htbp]
\centering
\begin{tabular}{llll}
\hline
 Parameters& Optimal values & Parameters &Optimal values
 \\
\hline
\hline
$f_1$ & $-4.3890$ & $f_3$&  $-9.9523$ \\
\hline
$f_4$ & $6.2262$&  $f_5$& $1.4250 $\\
\hline
$\lambda_1$ & $5.1656 \times 10^{-29}$ & $\lambda_2$ & $5.9774 \times 10^{-29}$\\
\hline
$u_1$ & $2.4915\times 10^{13}\ eV$ & $u$ & $4.3497 \times 10^{13}\ eV$\\
\hline
$H$ & $142.028\ GeV$ & $\chi$ & $100\ GeV$\\
\hline
$\Lambda$ & $2.4300\times 10^{15}\ eV$ &  & \\
\hline
\end{tabular}
\caption{The optimal values of Model's parameters for inverted mass hierarchy.
\label{tab:optimal values for inverted parameters}}
\end{table}
\section{Effective Majorana neutrino mass parameter}
\label{sec:Effec}
\begin{figure}[htbp]
\centering
\includegraphics[width=0.6\textwidth]{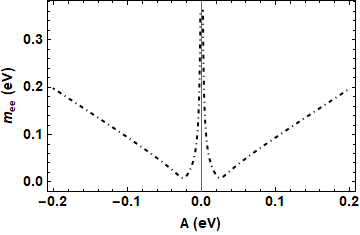}
\caption{Effective Majorana mass $\langle m_{ee} \rangle$ as a function of the model parameter $A$ in the case of normal mass hierarchy.}
\label{fig:mee_normal}
\end{figure}

\begin{figure}[htbp]
\centering
\includegraphics[width=0.6\textwidth]{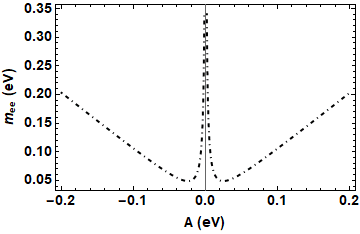}
\caption{Effective Majorana mass $\langle m_{ee} \rangle$ versus the parameter $A$ for the inverted mass hierarchy.}
\label{fig:mee_inverted}
\end{figure}

In the following section, we begin to find the effective Majorana neutrino mass parameter for normal as well as inverted mass hierarchy. This parameter is directly correlated to the amplitude of neutrinoless double beta $(0\nu\beta\beta)$ decay and is given as
\begin{equation}
\label{eq:e91}
\begin{aligned}
\left\langle m_{\text{ee}} \right\rangle = \left| \sum_{i = 1}^{3}{U_{\text{ei}}^{2}m_{i}} \right|
\end{aligned}
\end{equation}
where $U_{ei}^2$ denotes the components of squared $U_{PMNS}$ and $m_i$ denotes the masses of the three neutrinos. 
The effective neutrino masses associated with neutrinoless double beta decay ($\langle m_{ee} \rangle$) is calculated using the parameters obtained in section \ref{sec:Numericsal Analysis} for both of the neutrino mass hierarchy that yields the following results:
\begin{equation}
\label{effective majorana parametervalues}
\begin{aligned}
\langle {m_{ee}} \rangle=\begin{cases} 21.4051\ meV,
~~ &\text{in normal hierarchy (NH)} \\
50.3654\ meV,~~ &\text{in inverted hierarchy (IH)}
\end{cases}
\end{aligned}
\end{equation}
After imposing the $S_2$ symmetry, we get the expression for the effective Majorana neutrino mass parameter in the normal neutrino mass ordering by using the equations marked \ref{eq:e79}, as shown in equation \ref{eq:e92}.
\begin{equation}
\label{eq:e92}
\begin{aligned}
\langle m_{ee} \rangle = &\sqrt{A^2+\frac{3.9470\times10^{-7}}{A^2}-1.1837\times10^{-3}}
\end{aligned}
\end{equation}
Similarly, we get the expression for the effective Majorana neutrino mass parameter in the inverted neutrino mass ordering by using the equations marked \ref{eq:e86} and \ref{eq:e69}, as shown in equation \ref{eq:e93}.
\begin{equation}
\label{eq:e93}
\begin{aligned}
\langle m_{ee} \rangle=& \sqrt{ A^2+\frac{3.7061\times10^{-7}}{A^2}+1.1961\times10^{-3}}
\end{aligned}
\end{equation}
The variations of $\langle m_{ee} \rangle$ with the value of $A$ are shown in figures \ref{fig:mee_normal} and \ref{fig:mee_inverted}.

The following explicit values for $\langle m_{ee} \rangle$ are given for both normal as well as inverted mass hierarchy
\begin{equation}
\label{eq:e94}
\begin{aligned}
\langle m_{ee}\rangle = \begin{cases}
13.1448\ meV,~~ &\text{in normal hierarchy (NH)} \\
12.6608\ meV,~~ &\text{in inverted hierarchy (IH)}
\end{cases}
\end{aligned}
\end{equation}
As indicated by equation \ref{effective majorana parametervalues} and \ref{eq:e94}, the effective Majorana neutrino mass parameters 
associated with both normal and inverted neutrino mass orderings lie beyond 
the reach of current and forthcoming experiments focusing on neutrinoless 
double-beta decay $(0\nu\beta\beta)$. It's worth
noting that the upper limit for the Majorana 
neutrino mass parameter, denoted as $m_{ee}$, is 
constrained to be $\leq 140-380$ $meV$, and corresponds 
to a half-life of $T_{1/2}^{0\nu\beta\beta}
(_{}^{136} Xe) \geq 1.6 \times10^{25}$ 
years at a 
$90\%$ confidence level, as estimated by 
the EXO-200 experiment 
\cite{auger2012search}. This limit is 
expected to be refined in the near 
future. anticipations are high for the 
GERDA 
"phase-II" experiment 
\cite{agostinigerda}, which is 
projected 
to achieve a half-life of 
$T_{1/2}^{0\nu\beta\beta}(_{}^{76} 
Ge)\geq 2\times 10^{26}$ years, 
corresponding to a 
$\langle m_{ee} \rangle$ value of $\leq 100$ $meV$.
Concurrently, there is ongoing 
construction of a bolometric CUORE 
experiment employing $_{}^{130} Te$  
\cite{alessandria2011sensitivity}, with 
an estimated sensitivity of 
approximately $T_{1/2}^{0\nu\beta\beta}
(_{}^{130} Te) \sim10^{26}$ years, 
corresponding to $\langle m_{ee} \rangle \leq 50$ $meV$. 
In addition to these endeavors, 
ambitious plans 
are in place for next-generation 
$0\nu\beta\beta$ experiments utilizing 
ton-scale 
quantities of $_{}^{136}Xe$ \cite{auger2012search,agostinigerda} and
$_{}^{76}Ge$ 
\cite{aalseth2018search}.These projects 
aim to achieve sensitivities on the 
order of $T_{1/2}^{0\nu\beta\beta} 
\sim10^{27}$ years, 
corresponding to $\langle m_{ee}\rangle$ values in the 
range of approximately $12$ $meV$ to $30$ $meV$. It is noteworthy that our model 
predicts $T_{1/2}^{0\nu\beta\beta}$ 
values that align with the 
sensitivities expected from the 
upcoming generation or next-to-next 
generation $0\nu\beta\beta$ experiments.

\section{Scalar potential minimum 
condition}
\label{sec:Scalar}
The invariant scalar potential under 
$T_7$ is given as 
\begin{equation}
\label{eq:x1}
\begin{aligned}
V_{\text{total}} &= 
\mu_{\xi_{1}}^{2}\left( \xi_{1}\xi_{1} 
\right)_{1_{0}} + 
\lambda^{\xi_{1}}\left( \xi_{1}\xi_{1} 
\right)_{1_{0}}\left( \xi_{1}\xi_{1} 
\right)_{1_{0}} + 
\mu_{H\phi_{\nu}}^{2}\left( H\phi_{\nu} 
\right)_{1_{0}}\\ &+ 
\lambda_{1}^{H\phi_{\nu}}\left( 
H\phi_{\nu} \right)_{1_{0}}\left( 
H\phi_{\nu} \right)_{1_{0}} + 
\lambda_{2}^{H\phi_{\nu}}\left( 
H\phi_{\nu} \right)_{1_{1}}\left( 
H\phi_{\nu} \right)_{1_{2}}\\ &+ 
\lambda_{3}^{H\phi_{\nu}}\left( 
H\phi_{\nu} \right)_{3}\left( 
H\phi_{\nu} \right)_{\overline{3}} + 
\mu_{\phi_{l}\phi_{\nu}}^{2}\left( 
\phi_{l}\phi_{\nu} \right)_{1_{0}} \\&+
\lambda_{1}^{\phi_{l}\phi_{\nu}}\left( 
\phi_{l}\phi_{\nu} 
\right)_{1_{0}}\left( 
\phi_{l}\phi_{\nu} \right)_{1_{0}} + 
\lambda_{2}^{\phi_{l}\phi_{\nu}}\left( 
\phi_{l}\phi_{\nu} 
\right)_{1_{1}}\left( 
\phi_{l}\phi_{\nu} \right)_{1_{2}} \\&+ 
\lambda_{3}^{\phi_{l}\phi_{\nu}}\left( 
\phi_{l}\phi_{\nu} \right)_{3}\left( 
\phi_{l}\phi_{\nu} 
\right)_{\overline{3}} + 
\mu_{\xi_{2}\xi_{3}}^{2}\left( 
\xi_{3}\xi_{2} \right)_{1_{0}} + 
\lambda_{1}^{\xi_{2}\xi_{3}}\left( 
\xi_{3}\xi_{2} \right)_{1_{0}}\left( 
\xi_{3}\xi_{2} \right)_{1_{0}}\\ &+ 
\lambda_{1}^{\left( H\phi_{\nu} \right)
(\phi_{l}\phi_{\nu})}{(H\phi_{\nu})}_{3}
{(\phi_{l}\phi_{\nu})}_{\overline{3}} +
\lambda_{2}^{\left( H\phi_{\nu} \right)
(\phi_{l}\phi_{\nu})}
{(\phi_{l}\phi_{\nu})}_{3}
{(H\phi_{\nu})}_{\overline{3}} .
\end{aligned}
\end{equation}
After using the minimization conditions,
\begin{equation}
\label{eq:y}
\begin{aligned}
\frac{\partial V_{\text{total}}}{\partial\vartheta} = 0,\quad \frac{\partial^{2}V_{\text{total}}}{\partial\vartheta^{2}} > 0,
\end{aligned}
\end{equation}
where, $\vartheta =$ $h_{1}$, $h_{2}$, $h_{3}$, $\phi_{l_{1}}$, $\phi_{l_{2}}$, $\phi_{l_{3}}$, $\phi_{\nu_{1}}$, $\phi_{\nu_{2}}$, $\phi_{\nu_{3}}$, $\xi_{1}$, $\xi_{2}$, $\xi_{3}$, one can get following relations,
\begin{equation}
\label{eq:z1}
\begin{aligned}
\chi_{1}(\nu_{2}&\chi_{2}\lambda_{2}^{( 
H\phi_{\nu} )( \phi_{l}\phi_{\nu} 
)} + 2H_{1}\chi_{1}( 
\lambda_{1}^{H\phi_{\nu}} + 
\lambda_{2}^{H\phi_{\nu}} ) + ( 
H_{2}\chi_{2} + H_{3}\chi_{3} )( 
2\lambda_{1}^{H\phi_{\nu}} - 
\lambda_{2}^{H\phi_{\nu}}\\& + 
\lambda_{3}^{H\phi_{\nu}})  + 
\nu_{3}\chi_{3}\lambda_{1}^{( H\phi_{\nu} 
)( \phi_{l}\phi_{\nu} )} + 
\mu_{H\phi_{\nu}}^{2} ) = 0,
\end{aligned}
\end{equation}
\begin{equation}
\label{eq:z2}
\begin{aligned}
\chi_{2}(2H_{2}&\chi_{2}( \lambda_{1}^{H\phi_{\nu}} + \lambda_{2}^{H\phi_{\nu}} ) + H_{1}\chi_{1}( 2\lambda_{1}^{H\phi_{\nu}} - \lambda_{2}^{H\phi_{\nu}} + \lambda_{3}^{H\phi_{\nu}} ) + \chi_{3}( \nu_{3}\lambda_{2}^{( H\phi_{\nu} )( \phi_{l}\phi_{\nu} )}\\& + H_{3}( 2\lambda_{1}^{H\phi_{\nu}} - \lambda_{2}^{H\phi_{\nu}} + \lambda_{3}^{H\phi_{\nu}}) ) + \nu_{1}\chi_{1}\lambda_{1}^{( H\phi_{\nu} )( \phi_{l}\phi_{\nu} )} + \mu_{H\phi_{\nu}}^{2}) = 0,
\end{aligned}
\end{equation}
\begin{equation}
\label{eq:z3}
\begin{aligned}
\chi_{3}(\nu_{1}&\chi_{1}\lambda_{2}^{( H\phi_{\nu} )( \phi_{l}\phi_{\nu} )} + 2H_{2}\chi_{2}\lambda_{1}^{H\phi_{\nu}} + 2H_{3}\chi_{3}\lambda_{1}^{H\phi_{\nu}} - H_{2}\chi_{2}\lambda_{2}^{H\phi_{\nu}} + 2H_{3}\chi_{3}\lambda_{2}^{H\phi_{\nu}} \\&+ H_{2}\chi_{2}\lambda_{3}^{H\phi_{\nu}} + H_{1}\chi_{1}(2\lambda_{1}^{H\phi_{\nu}} - \lambda_{2}^{H\phi_{\nu}} + \lambda_{3}^{H\phi_{\nu}} ) + \nu_{2}\chi_{2}\lambda_{1}^{( H\phi_{\nu})( \phi_{l}\phi_{\nu})} + \mu_{H\phi_{\nu}}^{2}) = 0,
\end{aligned}
\end{equation}
\begin{equation}
\label{eq:z4}
\begin{aligned}
\chi_{1}&(H_{3} \chi_{3}\lambda_{2}^{( H\phi_{\nu})(\phi_{l}\phi_{\nu})} + 2\nu_{1}\chi_{1}( \lambda_{1}^{\phi_{l}\phi_{\nu}} + \lambda_{2}^{\phi_{l}\phi_{\nu}}) +( \nu_{2}\chi_{2} + \nu_{3}\chi_{3})( 2\lambda_{1}^{\phi_{l}\phi_{\nu}} - \lambda_{2}^{\phi_{l}\phi_{\nu}} \\&+ \lambda_{3}^{\phi_{l}\phi_{\nu}}) + H_{2}\chi_{2}\lambda_{1}^{\left( H\phi_{\nu} \right)( \phi_{l}\phi_{\nu})} + \mu_{\phi_{l}\phi_{\nu}}^{2}) = 0,
\end{aligned}
\end{equation}

\begin{equation}
\label{eq:z5}
\begin{aligned}
\chi_{2}&( H_{1}\chi_{1}\lambda_{2}^{(H\phi_{\nu})( \phi_{l}\phi_{\nu})} + 2\nu_{2}\chi_{2}\lambda_{1}^{\phi_{l}\phi_{\nu}} + 2\nu_{3}\chi_{3}\lambda_{1}^{\phi_{l}\phi_{\nu}} + 2\nu_{2}\chi_{2}\lambda_{2}^{\phi_{l}\phi_{\nu}} - \nu_{3}\chi_{3}\lambda_{2}^{\phi_{l}\phi_{\nu}} \\&+ \nu_{3}\chi_{3}\lambda_{3}^{\phi_{l}\phi_{\nu}} + \nu_{1}\chi_{1}(2\lambda_{1}^{\phi_{l}\phi_{\nu}} - \lambda_{2}^{\phi_{l}\phi_{\nu}} + \lambda_{3}^{\phi_{l}\phi_{\nu}}) + H_{3}\chi_{3}\lambda_{1}^{( H\phi_{\nu})( \phi_{l}\phi_{\nu})} + \mu_{\phi_{l}\phi_{\nu}}^{2}) = 0,
\end{aligned}
\end{equation}
\begin{equation}
\label{eq:z6}
\begin{aligned}
\chi_{3}&(H_{2}\chi_{2}\lambda_{2}^{( H\phi_{\nu})(\phi_{l}\phi_{\nu})} + 2\nu_{3}\chi_{3}( \lambda_{1}^{\phi_{l}\phi_{\nu}} + \lambda_{2}^{\phi_{l}\phi_{\nu}}) + \nu_{1}\chi_{1}( 2\lambda_{1}^{\phi_{l}\phi_{\nu}} - \lambda_{2}^{\phi_{l}\phi_{\nu}} + \lambda_{3}^{\phi_{l}\phi_{\nu}}) \\&+ \nu_{2}\chi_{2}(2\lambda_{1}^{\phi_{l}\phi_{\nu}} - \lambda_{2}^{\phi_{l}\phi_{\nu}} + \lambda_{3}^{\phi_{l}\phi_{\nu}}) + H_{1}\chi_{1}\lambda_{1}^{(H\phi_{\nu})( \phi_{l}\phi_{\nu})}+\mu_{\phi_{l}\phi_{\nu}}^{2}) = 0,
\end{aligned}
\end{equation}
\begin{equation}
\label{eq:z7}
\begin{aligned}
2{H_{1}}^{2}&\chi_{1}(\lambda_{1}^{H\phi_{\nu}} + \lambda_{2}^{H\phi_{\nu}}) + H_{1}( \nu_{2}\chi_{2}\lambda_{2}^{(H\phi_{\nu})( \phi_{l}\phi_{\nu})}+(H_{2}\chi_{2} + H_{3}\chi_{3} )(2\lambda_{1}^{H\phi_{\nu}} - \lambda_{2}^{H\phi_{\nu}} + \\&\lambda_{3}^{H\phi_{\nu}} ) + \nu_{3}\chi_{3}\lambda_{1}^{( H\phi_{\nu})( \phi_{l}\phi_{\nu})} + \mu_{H\phi_{\nu}}^{2}) + \nu_{1}( H_{3}\chi_{3}\lambda_{2}^{( H\phi_{\nu})(\phi_{l}\phi_{\nu})} + 2\nu_{1}\chi_{1}( \lambda_{1}^{\phi_{l}\phi_{\nu}} + \lambda_{2}^{\phi_{l}\phi_{\nu}} ) \\&+(\nu_{2}\chi_{2} + \nu_{3}\chi_{3})(2\lambda_{1}^{\phi_{l}\phi_{\nu}} - \lambda_{2}^{\phi_{l}\phi_{\nu}} + \lambda_{3}^{\phi_{l}\phi_{\nu}}) + H_{2}\chi_{2}\lambda_{1}^{( H\phi_{\nu})( \phi_{l}\phi_{\nu})} + \mu_{\phi_{l}\phi_{\nu}}^{2} ) = 0,
\end{aligned}
\end{equation}
\begin{equation}
\label{eq:z8}
\begin{aligned}
2{H_{2}}^{2}&\chi_{2}(\lambda_{1}^{H\phi_{\nu}} + \lambda_{2}^{H\phi_{\nu}}) + H_{1}\chi_{1}( \nu_{2}\lambda_{2}^{( H\phi_{\nu})( \phi_{l}\phi_{\nu})} + H_{2}( 2\lambda_{1}^{H\phi_{\nu}} - \lambda_{2}^{H\phi_{\nu}} + \lambda_{3}^{H\phi_{\nu}})) \\& + H_{2}( \nu_{3}\chi_{3}\lambda_{2}^{(H\phi_{\nu})( \phi_{l}\phi_{\nu})} + H_{3}\chi_{3}( 2\lambda_{1}^{H\phi_{\nu}} - \lambda_{2}^{H\phi_{\nu}} +\lambda_{3}^{H\phi_{\nu}}) + \nu_{1}\chi_{1}\lambda_{1}^{( H\phi_{\nu})( \phi_{l}\phi_{\nu} )} \\&+ \mu_{H\phi_{\nu}}^{2}) + \nu_{2}( 2\nu_{2}\chi_{2}( \lambda_{1}^{\phi_{l}\phi_{\nu}} + \lambda_{2}^{\phi_{l}\phi_{\nu}}) + \nu_{1}\chi_{1}( 2\lambda_{1}^{\phi_{l}\phi_{\nu}} - \lambda_{2}^{\phi_{l}\phi_{\nu}} + \lambda_{3}^{\phi_{l}\phi_{\nu}}) \\&+ \nu_{3}\chi_{3}( 2\lambda_{1}^{\phi_{l}\phi_{\nu}} - \lambda_{2}^{\phi_{l}\phi_{\nu}} + \lambda_{3}^{\phi_{l}\phi_{\nu}}) + H_{3}\chi_{3}\lambda_{1}^{( H\phi_{\nu})( \phi_{l}\phi_{\nu})} + \mu_{\phi_{l}\phi_{\nu}}^{2}) = 0,
\end{aligned}
\end{equation}
\begin{equation}
\label{eq:z9}
\begin{aligned}
2{H_{3}}^{2}\chi_{3}&(\lambda_{1}^{H\phi_{\nu}} + \lambda_{2}^{H\phi_{\nu}}) + H_{3}( \nu_{1}\chi_{1}\lambda_{2}^{( H\phi_{\nu})( \phi_{l}\phi_{\nu})} +( H_{1}\chi_{1} + H_{2}\chi_{2})(2\lambda_{1}^{H\phi_{\nu}}\\& - \lambda_{2}^{H\phi_{\nu}} + \lambda_{3}^{H\phi_{\nu}}) + \nu_{2}\chi_{2}\lambda_{1}^{( H\phi_{\nu}( \phi_{l}\phi_{\nu})} + \mu_{H\phi_{\nu}}^{2}) + \nu_{3}( H_{2}\chi_{2}\lambda_{2}^{( H\phi_{\nu})( \phi_{l}\phi_{\nu})} \\& + 2\nu_{3}\chi_{3}( \lambda_{1}^{\phi_{l}\phi_{\nu}} + \lambda_{2}^{\phi_{l}\phi_{\nu}}) + \nu_{1}\chi_{1}( 2\lambda_{1}^{\phi_{l}\phi_{\nu}} - \lambda_{2}^{\phi_{l}\phi_{\nu}} + \lambda_{3}^{\phi_{l}\phi_{\nu}}) \\&+ \nu_{2}\chi_{2}( 2\lambda_{1}^{\phi_{l}\phi_{\nu}} - \lambda_{2}^{\phi_{l}\phi_{\nu}} + \lambda_{3}^{\phi_{l}\phi_{\nu}}) + H_{1}\chi_{1}\lambda_{1}^{( H\phi_{\nu})( \phi_{l}\phi_{\nu})} + \mu_{\phi_{l}\phi_{\nu}}^{2}) = 0,
\end{aligned}
\end{equation}

\begin{equation}
\label{eq:z10}
\begin{aligned}
2u_{1}\left( 2{u_{1}}^{2}\lambda^{\xi_{1}} + \mu_{\xi_{1}}^{2} \right) = 0,
\end{aligned}
\end{equation}
\begin{equation}
\label{eq:z}
\begin{aligned}
u_{3}\left( 2u_{2}u_{3}\lambda_{1}^{\xi_{2}\xi_{3}} + \mu_{\xi_{2}\xi_{3}}^{2} \right) = 0,
\end{aligned}
\end{equation}
\begin{equation}
\label{eq:z11}
\begin{aligned}
u_{2}\left( 2u_{2}u_{3}\lambda_{1}^{\xi_{2}\xi_{3}} + \mu_{\xi_{2}\xi_{3}}^{2} \right) = 0,
\end{aligned}
\end{equation}
and
\begin{equation}
\label{eq:z12}
\begin{aligned}
2{\chi_{1}}^{2}\left( \lambda_{1}^{H\phi_{\nu}} + \lambda_{2}^{H\phi_{\nu}} \right) > 0,
\end{aligned}
\end{equation}
\begin{equation}
\label{eq:z13}
\begin{aligned}
2{\chi_{2}}^{2}\left( \lambda_{1}^{H\phi_{\nu}} + \lambda_{2}^{H\phi_{\nu}} \right) > 0,
\end{aligned}
\end{equation}
\begin{equation}
\label{eq:z14}
\begin{aligned}
2{\chi_{3}}^{2}\left( \lambda_{1}^{H\phi_{\nu}} + \lambda_{2}^{H\phi_{\nu}} \right) > 0,
\end{aligned}
\end{equation}
\begin{equation}
\label{eq:z15}
\begin{aligned}
2{\chi_{1}}^{2}\left( \lambda_{1}^{\phi_{l}\phi_{\nu}} + \lambda_{2}^{\phi_{l}\phi_{\nu}} \right) > 0,
\end{aligned}
\end{equation}
\begin{equation}
\label{eq:z16}
\begin{aligned}
2{\chi_{2}}^{2}\left( \lambda_{1}^{\phi_{l}\phi_{\nu}} + \lambda_{2}^{\phi_{l}\phi_{\nu}} \right) > 0,
\end{aligned}
\end{equation}
\begin{equation}
\label{eq:z17}
\begin{aligned}
2{\chi_{3}}^{2}\left( \lambda_{1}^{\phi_{l}\phi_{\nu}} + \lambda_{2}^{\phi_{l}\phi_{\nu}} \right) > 0,
\end{aligned}
\end{equation}
\begin{equation}
\label{eq:z18}
\begin{aligned}
2{H_{1}}^{2}\left( \lambda_{1}^{H\phi_{\nu}} + \lambda_{2}^{H\phi_{\nu}} \right) + 2{\nu_{1}}^{2}\left( \lambda_{1}^{\phi_{l}\phi_{\nu}} + \lambda_{2}^{\phi_{l}\phi_{\nu}} \right) > 0,
\end{aligned}
\end{equation}
\begin{equation}
\label{eq:z19}
\begin{aligned}
2{H_{2}}^{2}\left( \lambda_{1}^{H\phi_{\nu}} + \lambda_{2}^{H\phi_{\nu}} \right) + 2{\nu_{2}}^{2}\left( \lambda_{1}^{\phi_{l}\phi_{\nu}} + \lambda_{2}^{\phi_{l}\phi_{\nu}} \right) > 0,
\end{aligned}
\end{equation}
\begin{equation}
\label{eq:z20}
\begin{aligned}
2{H_{3}}^{2}\left( \lambda_{1}^{H\phi_{\nu}} + \lambda_{2}^{H\phi_{\nu}} \right) + 2{\nu_{3}}^{2}\left( \lambda_{1}^{\phi_{l}\phi_{\nu}} + \lambda_{2}^{\phi_{l}\phi_{\nu}} \right) > 0,
\end{aligned}
\end{equation}
\begin{equation}
\label{eq:z21}
\begin{aligned}
2\left( 6{u_{1}}^{2}\lambda^{\xi_{1}} + \mu_{\xi_{1}}^{2} \right) > 0,
\end{aligned}
\end{equation}
\begin{equation}
\label{eq:z22}
\begin{aligned}
2{u_{3}}^{2}\lambda_{1}^{\xi_{2}\xi_{3}} > 0,
\end{aligned}
\end{equation}
\begin{equation}
\label{eq:z23}
\begin{aligned}
2{u_{2}}^{2}\lambda_{1}^{\xi_{2}\xi_{3}} > 0.
\end{aligned}
\end{equation}
The minimization requirements of this potential lead to the extremum solutions listed below,
\begin{equation}
\label{eq:vacuum}
\begin{aligned}
\left\langle h \right\rangle = (H,H,&H),\quad \left\langle \phi_{l} \right\rangle = \left( \nu_{1},\nu_{2},\nu_{3} \right),\quad 
\left\langle \phi_{\nu} \right\rangle = 
(\chi,\chi,\chi),\\ &\langle \xi_{1} \rangle 
= u_{1},\quad \left\langle \xi_{2} \right\rangle = u,\quad 
\left\langle \xi_{3} \right\rangle = u,
\end{aligned}
\end{equation}
with the conditions
\begin{equation}
\label{eq:z24}
\begin{aligned}
\left( \lambda_{3}^{H\phi_{\nu}}\left| \lambda_{3}^{\phi_{l}\phi_{\nu}} \right|\lambda_{1}^{(H\phi_{\nu})(\phi_{l}\phi_{\nu})} \middle| \mu_{H\phi_{\nu}} \right) \in \text{Reals},
\end{aligned}
\end{equation}
\begin{equation}
\label{eq:z25}
\begin{aligned}
\lambda^{\xi_{1}} > 0,
\end{aligned}
\end{equation}
\begin{equation}
\label{eq:z26}
\begin{aligned}
\lambda_{1}^{\xi_{2}\xi_{3}} > 0,
\end{aligned}
\end{equation}
\begin{equation}
\label{eq:z27}
\begin{aligned}
\lambda_{1}^{\phi_{l}\phi_{\nu}} < - \frac{\lambda_{3}^{\phi_{l}\phi_{\nu}}}{3},
\end{aligned}
\end{equation}
\begin{equation}
\label{eq:z28}
\begin{aligned}
\lambda_{2}^{(H\phi_{\nu})(\phi_{l}\phi_{\nu})} < - \lambda_{1}^{(H\phi_{\nu})(\phi_{l}\phi_{\nu})},
\end{aligned}
\end{equation}
\begin{equation}
\label{eq:z29}
\begin{aligned}
\mu_{\phi_{l}\phi_{\nu}} < 0,
\end{aligned}
\end{equation}
\begin{equation}
\label{eq:z30}
\begin{aligned}
\mu_{H\phi_{\nu}} < - \sqrt{\frac{\lambda_{2}^{(H\phi_{\nu})(\phi_{l}\phi_{\nu})}\mu_{\phi_{l}\phi_{\nu}}^{2} + \lambda_{1}^{(H\phi_{\nu})(\phi_{l}\phi_{\nu})}\mu_{\phi_{l}\phi_{\nu}}^{2}}{3\lambda_{1}^{\phi_{l}\phi_{\nu}} + \lambda_{3}^{\phi_{l}\phi_{\nu}}}},
\end{aligned}
\end{equation}
\begin{equation}
\label{eq:z31}
\begin{aligned}
- \frac{5\lambda_{3}^{H\phi_{\nu}}}{9} &< \lambda_{1}^{H\phi_{\nu}} <(- 9\lambda_{1}^{\phi_{l}\phi_{\nu}}\mu_{H\phi_{\nu}}^{4} - 3\lambda_{3}^{\phi_{l}\phi_{\nu}}\mu_{H\phi_{\nu}}^{4} + 3\lambda_{2}^{(H\phi_{\nu})(\phi_{l}\phi_{\nu})}\mu_{H\phi_{\nu}}^{2}\mu_{\phi_{l}\phi_{\nu}}^{2} \\&+ 3\lambda_{1}^{(H\phi_{\nu})(\phi_{l}\phi_{\nu})}\mu_{H\phi_{\nu}}^{2}\mu_{\phi_{l}\phi_{\nu}}^{2}- 5\lambda_{3}^{H\phi_{\nu}}\mu_{\phi_{l}\phi_{\nu}}^{4})/(9\mu_{\phi_{l}\phi_{\nu}}^{4}),
\end{aligned}
\end{equation}
\begin{equation}
\label{eq:z32}
\begin{aligned}
\lambda_{2}^{\phi_{l}\phi_{\nu}} > - \lambda_{1}^{\phi_{l}\phi_{\nu}}.
\end{aligned}
\end{equation}

The following low-energy scalar potential (a comparable approach is adopted in the work \cite{bonilla2023fermion}), allows us to focus on the low-energy phenomenology:
\begin{equation}
\label{eq:lowenergypotential}
\begin{aligned}
V&= 
\mu_{\xi_{1}}^{2}\left( \xi_{1}\xi_{1} 
\right)_{1_{0}} + 
\lambda^{\xi_{1}}\left( \xi_{1}\xi_{1} 
\right)_{1_{0}}\left( \xi_{1}\xi_{1} 
\right)_{1_{0}} + 
\mu_{H\phi_{\nu}}^{2}\left( H\phi_{\nu} 
\right)_{1_{0}}+ 
\lambda_{1}^{H\phi_{\nu}}\left( 
H\phi_{\nu} \right)_{1_{0}}\left( 
H\phi_{\nu} \right)_{1_{0}} \\&+ 
\lambda_{2}^{H\phi_{\nu}}\left( 
H\phi_{\nu} \right)_{1_{1}}\left( 
H\phi_{\nu} \right)_{1_{2}}+ 
\mu_{\phi_{l}\phi_{\nu}}^{2}\left( 
\phi_{l}\phi_{\nu} \right)_{1_{0}}+
\lambda_{1}^{\phi_{l}\phi_{\nu}}\left( 
\phi_{l}\phi_{\nu} 
\right)_{1_{0}}\left( 
\phi_{l}\phi_{\nu} \right)_{1_{0}} \\&+ 
\lambda_{2}^{\phi_{l}\phi_{\nu}}\left( 
\phi_{l}\phi_{\nu} 
\right)_{1_{1}}\left( 
\phi_{l}\phi_{\nu} \right)_{1_{2}} + 
\mu_{\xi_{2}\xi_{3}}^{2}\left( 
\xi_{3}\xi_{2} \right)_{1_{0}} + 
\lambda_{1}^{\xi_{2}\xi_{3}}\left( 
\xi_{3}\xi_{2} \right)_{1_{0}}\left( 
\xi_{3}\xi_{2} \right)_{1_{0}}.
\end{aligned}
\end{equation}
By minimizing this scalar potential, we derive these effective mass terms.
\begin{equation}
\label{eq:lowenermasses1}
\begin{aligned}
{\mu^2_{\xi_1}=- 2u_{1}^{2}\lambda^{\xi}_{1}}
\end{aligned}
\end{equation}
\begin{equation}
\label{eq:lowenermasses2}
\begin{aligned}
\mu^2_{H {\phi}_{\nu}}=-6 H\chi \lambda^{H {\phi}_{\nu}}_{1}
\end{aligned}
\end{equation}
\begin{equation}
\label{eq:lowenermasses3}
\begin{aligned}
\mu^{2}_{\xi_{2}\xi_{3}}=-2u^{2}\lambda_{1}^{\xi_{2}\xi_{3}}
\end{aligned}
\end{equation}
\section{Conclusion}
\label{sec:Conclusion}
The model presented in this paper is constructed based on the $T_7$ discrete flavor symmetry. The model effectively incorporates the masses and mixings of fermions. In this model, the light neutrino mass matrix is derived using seesaw mechanism of type I, which involves the Dirac neutrino mass matrix as well as the right-handed neutrino mass matrix. In this work, we employed the particle swarm optimization (PSO) algorithm to determine the optimal set of parameters for computing the PMNS mixing matrix ($U_{\text{PMNS}}$) and neutrino mass observables. For the normal mass hierarchy (NH), the predicted neutrino masses are $m_1 = 19.998\ \text{meV}$, $m_2 = 21.7904 \ \text{meV}$, and $m_3 = 53.9715\ \text{meV}$, while for the inverted hierarchy (IH), the values are $m_1 = 51.0882\ \text{meV}$, $m_2 = 51.816\ \text{meV}$, and $m_3 = 10\ \text{meV}$. The effective Majorana neutrino mass parameter, are also computed as $\langle m_{ee} \rangle = 21.4051\ \text{meV}$ (NH) and $\langle m_{ee} \rangle = 50.3654\ \text{meV}$ (IH), respectively. The PSO method is found to be effective in fitting the model parameters for the two different NH and IH cases, with the resultant predictions agreeing with existing global neutrino oscillation data. The model is also testable through a prediction for the effective Majorana mass parameter $\langle m_{ee} \rangle$ for experiments for neutrinoless double beta decay. The model is consistent with the most up-to-date cosmological limits for the total neutrino mass sum, $\sum m_i \leq 0.12$ eV. Furthermore, the model is able to predict the mixing angles $\theta_{12}, \theta_{13}, \theta_{23}$, the Dirac $CP$-violating phase $\delta$, and the two Majorana phases $\beta_1$ and $\beta_2$ with all predictions within the range allowed for experimentation, rendering the model phenomenologically consistent and experimentally accessible to future searches. We also analyzed the construction of the $T_7$ flavor model by further imposing an additional $S_2$ flavor symmetry. Due to this dilution, the agreement between the measured values of leptonic mixing angles and their experimental values leads to the emergence of a non-zero Dirac $CP$-violating phase in the realm of leptons, which equals $-\pi/2$. After imposing $S_2$ symmetry model predicts mixing angles as $\theta_{13}=7.9315^\circ$, 
$\theta_{23}=44.9980^\circ$ and
$\theta_{12}=35.6320^\circ$, which is a remarkable deviation from maximal mixing. The masses of the three neutrinos are estimated as $m_1=10\ meV$, 
$m_2=13.225\ meV$ and $m_3=51.118 \ meV$ for normal hierarchy, $m_1=50.350\ meV$, $m_2=51.088\ meV$ and $m_3=10\ meV$, for inverted hierarchy. Furthermore, the model predicts an effective Majorana neutrino mass parameter relevant for neutrinoless double beta decay, with values $\langle m_{ee} \rangle = 13.1448~\text{meV}$ ($12.6608~\text{meV}$) for the normal (inverted) neutrino mass hierarchy.
\appendix
\section{\texorpdfstring{$T_7$}{TEXT} group and its multiplication rules}
\label{sec:App}
The group $T_7$ \cite{bookfordiscretesymmetries} is a 
subgroup of  $SU(3)$, possessing $21$ elements and is isomorphic
to ${Z_7} \times {Z_3}$. It is also called Frobenius group. $T_7$ group contains three singlet irreducible representations $\textbf{1}_k$ here $k=0, 1, 2$  and two triplets $\textbf{3}$ and $\overline{\textbf{3}}$.

The tensor products between triplets are
\begin{equation}
\label{eq:ap1}
\begin{aligned}
\begin{pmatrix}
a_{1} \\
a_{2} \\
a_{4} \\
\end{pmatrix}_{\mathbf{3}} \otimes \begin{pmatrix}
b_{1} \\
b_{2} \\
b_{4} \\
\end{pmatrix}_{\mathbf{3}} = \begin{pmatrix}
a_{4}b_{4} \\
a_{1}b_{1} \\
a_{2}b_{2} \\
\end{pmatrix}_{\mathbf{3}} \oplus \begin{pmatrix}
a_{2}b_{4} \\
a_{4}b_{1} \\
a_{1}b_{2} \\
\end{pmatrix}_{\overline{\mathbf{3}}} \oplus \begin{pmatrix}
a_{4}b_{2} \\
a_{1}b_{4} \\
a_{2}b_{1} \\
\end{pmatrix}_{\overline{\mathbf{3}}}
\end{aligned}
\end{equation}
\begin{equation}
\label{eq:ap2}
\begin{aligned}
\begin{pmatrix}
a_{6} \\
a_{5} \\
a_{3} \\
\end{pmatrix}_{\overline{\mathbf{3}}} \otimes \begin{pmatrix}
b_{6} \\
b_{5} \\
b_{3} \\
\end{pmatrix}_{\overline{\mathbf{3}}} = \begin{pmatrix}
a_{5}b_{3} \\
a_{3}b_{6} \\
a_{6}b_{5} \\
\end{pmatrix}_{\mathbf{3}} \oplus \begin{pmatrix}
a_{3}b_{5} \\
a_{6}b_{3} \\
a_{5}b_{6} \\
\end{pmatrix}_{\mathbf{3}} \oplus \begin{pmatrix}
a_{3}b_{3} \\
a_{6}b_{6} \\
a_{5}b_{5} \\
\end{pmatrix}_{\overline{\mathbf{3}}}
\end{aligned}
\end{equation}
\begin{equation}
\label{eq:ap3}
\begin{aligned}
\begin{pmatrix}
a_{1} \\
a_{2} \\
a_{4} \\
\end{pmatrix}_{\mathbf{3}} \otimes \begin{pmatrix}
b_{6} \\
b_{5} \\
b_{3} \\
\end{pmatrix}_{\overline{\mathbf{3}}} = \begin{pmatrix}
a_{2}b_{6} \\
a_{4}b_{5} \\
a_{1}b_{3} \\
\end{pmatrix}_{\mathbf{3}} \oplus \quad \begin{pmatrix}
a_{1}b_{5} \\
a_{2}b_{3} \\
a_{4}b_{6} \\
\end{pmatrix}_{\overline{\mathbf{3}}}\oplus\sum_{k = 0,1,2}^{}\left( a_{1}b_{6} + \omega^{2k}a_{2}b_{5} + \omega^{k}a_{4}b_{3} \right)_{\mathbf{1}_{k}}
\end{aligned}
\end{equation}
Singlet carries the tensor product as
\begin{equation}
\label{eq:ap4}
\begin{aligned}
{(a)}_{\mathbf{1}_{0}} \otimes \ {(b)}_{\mathbf{1}_{0}} = \ {(a)}_{\mathbf{1}_{1}} \otimes \ {(b)}_{\mathbf{1}_{2}} = \ {(a)}_{\mathbf{1}_{2}} \otimes \ {(b)}_{\mathbf{1}_{1}} = {(ab)}_{\mathbf{1}_{0}}
\end{aligned}
\end{equation}
Singlet has no impact on triplet representations. The tensor product between triplets and singlets is as follows
\begin{equation}
\label{eq:ap5}
\begin{aligned}
\mathbf{3} \otimes \mathbf{1}_{k} = \mathbf{3},\qquad \overline{\mathbf{3}} \otimes \mathbf{1}_{k} = \overline{\mathbf{3}}
\end{aligned}
\end{equation}
\bibliographystyle{apsrev}
\bibliography{biblio}
\end{document}